\documentstyle[psfig]{l-aa}     

\def\ros{{\sl ROSAT }}

\def\ein{{\sl Einstein }}

\newcommand\approxlt{\mbox{$^{<}\hspace{-0.24cm}_{\sim}$}}
\def\asec{\ifmmode ^{\prime\prime}\else$^{\prime\prime}$\fi}
\def\amin{\ifmmode ^{\prime}\else$^{\prime}$\fi}

\def\it{\sl}

\begin{document}
           
   \thesaurus{06         
              (08.02.1;  
               11.13.1;  
               13.25.5;  
               08.09.2)  
             }
   \title{X-Ray Binary Systems in the Small Magellanic Cloud}
 
   \author{P. Kahabka\inst{1,2}, 
           W. Pietsch\inst{3}
          }

   \offprints{P. Kahabka}
 
   \institute{$^1$~Astronomical Institute, University of Amsterdam, 
              Kruislaan 403, NL--1098 SJ Amsterdam, The Netherlands\\
              $^2$~Center for High Energy Astrophysics, University of
              Amsterdam, Kruislaan 403, NL--1098 SJ Amsterdam, The
              Netherlands\\
              $^3$~Max-Planck-Institut f\"ur extraterrestrische Physik,
              D--85740 Garching bei M\"unchen, Federal Republic of Germany}

   \date{Received May 29, 1995; accepted February 14, 1996}
 
   \maketitle

   \begin{abstract}

We present the result of a systematic search for spectrally hard and soft
X-ray binary systems in the Small Magellanic Cloud (SMC). This search has
been applied to \ros {\sl PSPC} data (0.1-2.4~keV) collected during nine 
pointed observations towards this galaxy covering a time span of $\sim$2~years
from October~91 till October~93. Strict selection criteria have been defined 
in order to confine the sample of candidates. Finally seven spectrally hard 
and four spectrally soft sources were selected from the list as candidates 
for binaries in the SMC. The sample is luminosity limited (above $\rm 
\sim 3\times 10^{35}\ erg\ s^{-1}$). SMC~X-1 has been observed during a full 
binary orbit starting with a low-state covering an X-ray eclipse and emerging 
into a bright long-duration flare with two short-duration flares separated 
by $\rm \sim 10~hours$. The Be type transient SMC~X-2 has been redetected by 
\ros (second reported outburst). Variability has been found in the X-ray 
source RX~J0051.8-7231 already discovered with \ein and in RX~J0052.1-7319 
which is also known from \ein observations. RX~J0101.0-7206 has been 
discovered at the north-eastern boundary of the giant SMC HII region N66 
during an X-ray outburst and half a year later during a quiescent phase. A 
variable source, RX~J0049.1-7250, located north-east of the SMC supernova 
remnant N~19 and which may either be an X-ray binary or an AGN turns out to 
be strongly absorbed. It may be located behind the SMC. If it is an X-ray 
binary then it radiates at the Eddington limit in the X-ray bright state. 
Another variable and hard X-ray source RX~J0032.9-7348 has been discovered 
at the south-eastern border of the body of the SMC. A high mass X-ray binary
nature is favored for this source. 

A high mass X-ray binary nature is favored for the persistent sources 
where an optical counterpart of spectral type O or B has been identified. 
A possible Be type nature is favored for the few transient X-ray sources 
for which an optical identification with a B star has been achieved. We find
about equal numbers of persistent (and highly variable) and transient X-ray
binaries and binary candidates. Sources for which no optical candidate has 
been found in catalogs are candidate low-mass X-ray binaries (LMXBs) or 
black hole binaries. We searched for CAL~87 like systems in the SMC 
pointed catalog and found none. This implies that these systems are very 
rare and currently not existent in the SMC. A new candidate supersoft source
RX~J0103.8-7254 has been detected. We cannot exclude that it is a foreground 
object. 

%
 
   \end{abstract}

%

\section{Introduction}

The study of X-ray binary populations in galaxies outside of our own Galaxy 
is of major interest. First the evolution of our Galaxy is specific which 
reflects itself in their different stellar populations, their ages and 
distributions. The Magellanic Clouds (MCs), satellites of our own Galaxy, 
show different chemical compositions are irregular in shape and are heavily
interacting with our Galaxy (Gardiner et al. 1994). This influences the star 
formation history. Any kind of study of stellar populations is therefore of 
particular interest. X-ray binaries are interacting binaries with mass 
transfer from a donor star to a compact object, a white dwarf, neutron 
star or a black hole. They are evolved objects with the compact object 
being the product of evolution of a more massive star (cf. van den Heuvel 
1994, Livio 1994). The neutron star (NS) binaries are classified into high- 
and low-mass X-ray binaries (HMXBs and LMXBs). The HMXBs have donor stars 
more massive than $\sim$8-10~$\rm M_{\odot}$ and belong to the youngest 
stellar population of the Galaxy (age $\rm \approxlt 10^{7}\ yr$). The LMXBs 
belong to a much older stellar population (age $\rm 5-15\times10^{9}\ yr$). 
They are concentrated in the Galactic bulge and in the globular clusters 
(van den Heuvel 1992). 

The Magellanic System (MS) can be described as a binary system consisting 
of the LMC and the SMC which are orbiting around the Galaxy (cf. Gardiner 
et al. 1994).
In this description the MS forms a great circle in the sky perpendicular to
the Galactic poles. The MCs are presently near the perigalacticon and were
never closer to the Galaxy. Due to the frictional force in the Galactic
halo the orbital period of the MCs about the Galaxy is decaying (last orbital
period $\sim$1.3~Gyr, initial orbital period $\sim$3~Gyr). The present 
centre-to-centre distance between the MCs is $\sim$21~kpc. A close encounter
between the MCs occurred $\sim$0.2-0.4~Gyr ago at a distance of 7~kpc. The
previous encounter was at a much larger distance of 13~kpc. This may have
stimulated star formation in the MCs ($\sim$0.2~Gyr ago) and may explain why
the star formation histories in the MCs are so different from those in the
Galaxy. There is evidence of recent externally stimulated star formation in
the {\it Wing} of the SMC. The age of the stellar associations in the SMC
{\it Wing} and in the inter-Cloud region is less than 0.05~Gyr. This would
require a delay of at least 80~million years between the last LMC-SMC 
encounter and the onset of star formation activity. The spatial structure
of the SMC has e.g. been deduced from observations of Cepheids. According to
this view the SMC is composed of a 5-to-1 bar (seen edge-on), a near arm
in the north-east, and a far arm in the south-west. There is in addition 
material pulled out of the centre of the SMC (by the LMC) which is seen in
projection in front of the arm in the south-west (cf. Caldwell \& Coulson
1986). A point to note is that the SMC spiral arms are very young. It is not 
expected that the old population of LMXBs have yet formed there. Actually 
before \ros no single candidate LMXB has been known in the SMC. In the SMC 
bulge LMXBs may well have formed due to its considerable age of $\sim$15~Gyr. 
But possible kicks and small number statistics make it difficult to test 
these scenarii. 

Before the observations of the SMC with \ros six hard X-ray binary sources 
have been reported in the SMC (Clark et al. 1978, Seward \& Mitchell 1981, 
Inoue et al. 1983, Bruhweiler et al. 1987, Wang \& Wu 1992, Whitlock \& 
Lochner 1994). The first SMC X-ray sources discovered with rockets, 
{\it Uhuru}, {\it SAS-3} and HEAO-1 have been found to be superluminous with 
X-ray luminosities in excess of $\rm 10^{38} erg\ sec^{-1}$ and in part of 
transient nature (SMC~X-1, SMC~X-2, SMC~X-3 and H0107-750). The other two 
sources RX~J0051.8-7231 and RX~J0052.1-7319 have been first reported from 
{\it Einstein} observations. The \ros sources are not yet firmly identified 
with OB stars.

\ros performed several pointings to the field of the SMC (body and wing). A 
point source catalog of 250 (preliminary and unscreened) X-ray sources has 
been compiled from 9 pointed observations carried out by the authors of this
paper. The total list of sources detected in these fields will be presented 
in a separate paper. The X-ray sources have been classified into five 
categories. The new and distinct class of supersoft sources has been presented
in Kahabka, Pietsch \& Hasinger (1994). The new sample of supernova remnants 
detected in the SMC will be published in Kahabka et al. (1996). The bulk of 
sources, background AGNs and foreground stars, will be discussed together with
the total source list. 

We report in this paper about the hard and soft X-ray binary population 
studied with \ros in the SMC. In section~2 we describe the \ros observations, 
the data analysis and the selection criteria for hard and soft sources. In 
section~3 we report about the observational results of the individual systems,
the temporal and the spectral properties achieved during the pointed 
observations and results for SMC~X-1 during the all-sky survey observation. 
Then we detail results on the individual systems considering information 
deduced in previous work. In section~4 we investigate source variability and 
transient behavior, the luminosity distribution, extrapolate from the 
discussed SMC populations to the Galactic populations and discuss the nature 
of the rejected candidate X-ray binaries. Finally we summarize in section~5 
our results.  

%

\section{Observations and data analysis}
 
The observations reported in this paper were carried out with the {\it PSPC}
detector onboard the \ros observatory during the all-sky survey (RASS) 
and during 9 pointed observations in the time from 8~October 1991 to 
14~October~1993. The satellite, it's X-ray telescope (XRT) and the focal plane
detector ({\it PSPC}) used have been discussed in detail by Tr\"umper et al. 
(1983) and Pfeffermann et al. (1986).

\subsection{All-sky survey observations}

The field of the SMC covered by the pointed observations has been observed 
in the {\sl RASS} survey between 21~October 1990 and 31~October 1990. In the 
direction to the SMC the survey is complete to an apparent luminosity (not
corrected for interstellar absorption) of $\rm \sim 3\times 10^{35} erg\ 
s^{-1}$ (Kahabka \& Pietsch 1993). The corresponding intrinsic luminosity is 
$\rm \sim 8\times10^{35}\ erg\ s^{-1}$, as found from applying the constraints 
(conversion factor) from section~2.4 used for the pointed data.

\subsection{ROSAT pointed observations}

   \begin{figure*}
      \centering{
      \vbox{\psfig{figure=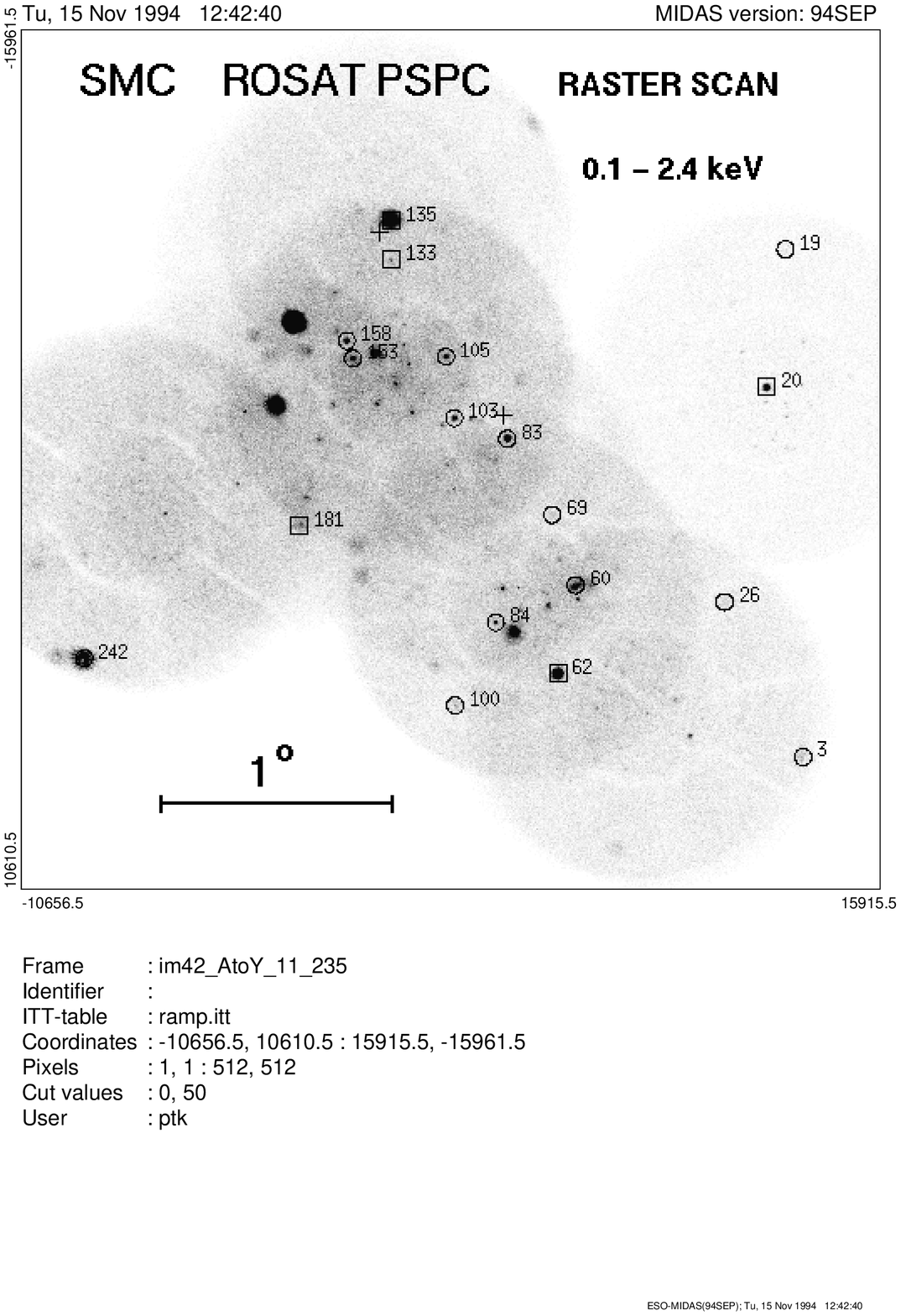,width=18.cm,%
       bbllx=3.05cm,bblly=10.55cm,bburx=18.8cm,bbury=26.1cm,clip=}}\par
       }
      \caption[]{Image (0.1-2.4~keV) of the SMC pointings A,B,C,D,E,F,G,X 
                 and Y (cf. Table~1). The center of the image is at R.A.(2000)
                 = $\rm 1^h\ 00^m$, Decl.(2000) = $\rm -73^d\ 00^m$. The 13 
                 candidate hard X-ray binaries in the SMC are marked with a 
                 circle and an internal catalog number is given (cf. Table~3).
                 The 5 candidate supersoft sources in the SMC are marked with
                 a square and an internal catalog number is given (cf. 
                 Table~4). The positions of the two transient X-ray binary 
                 sources SMC~X-3 (close to source \#83) and RX~J0059.2-7138 
                 (close to source \#135) which have not been detected in the 
                 analysed pointings are marked with a cross.}
         \label{FigGam}
    \end{figure*}

The pointed observations reported in this paper will be designated as 
regions A, B, C, D, E, F, G, X and Y. Table~1 gives a log of the field 
centers, observation intervals and exposure times. Regions A and B have 
originally been observed with a reduced time (observations A1 and B1) and 
about half a year later the observations were completed (observations A2 
and B2). A merged (0.1-2.4~keV) X-ray image of these fields only using the 
inner $\rm 45\amin$ of the detector is shown in Figure~1. The coverage due 
to the inner $\rm 20\amin$ of the detector for all pointings is shown in 
Figure~2.

\begin{figure}
      \centering{
      \vbox{\psfig{figure=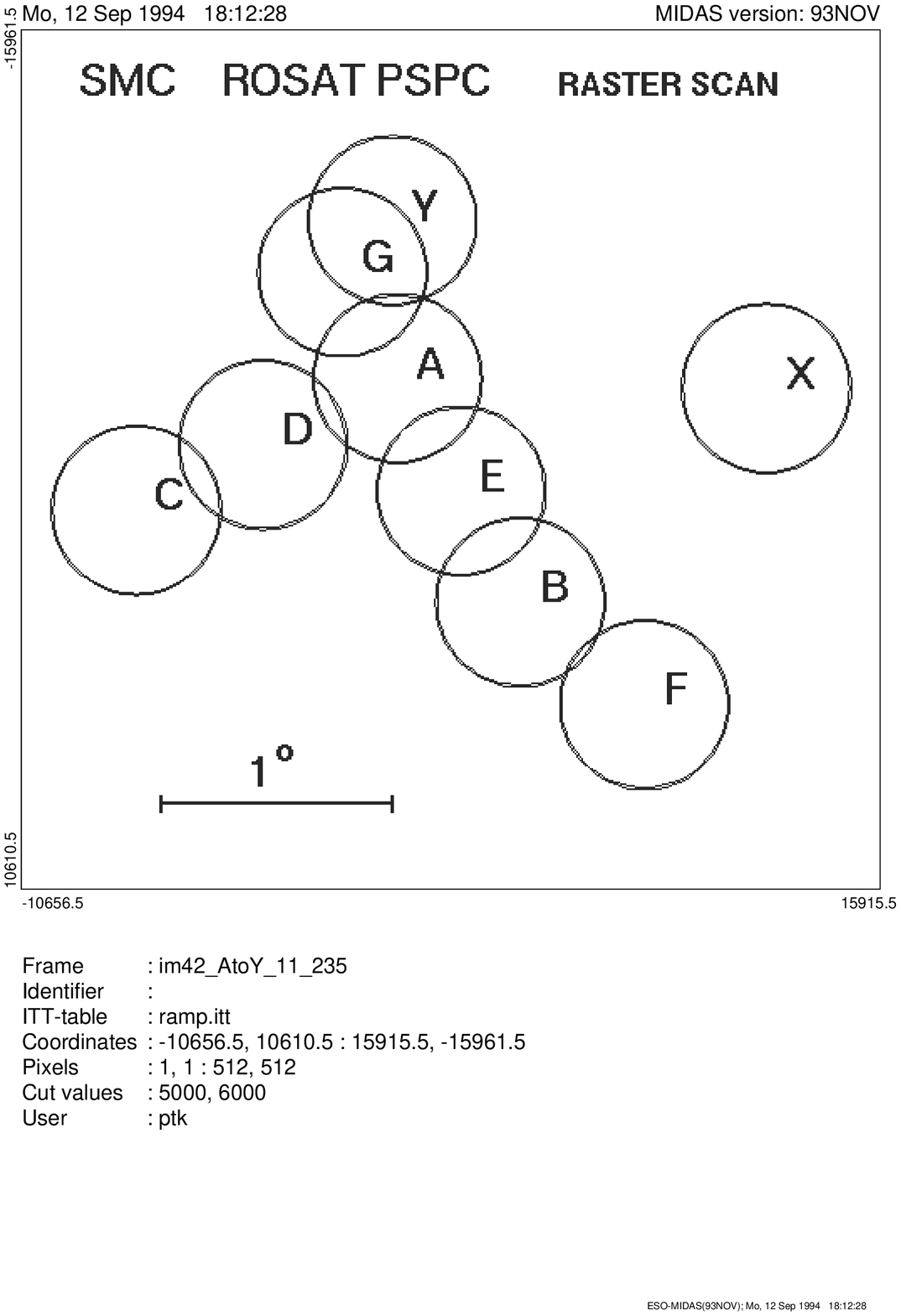,width=8.5cm,%
       bbllx=3.05cm,bblly=10.55cm,bburx=18.8cm,bbury=26.1cm,clip=}}\par
       }
      \caption[]{Image of the inner $\rm 20\amin$ of all pointings on the 
                 SMC discussed in the paper. The same field of view and 
                 projection is chosen as in Figure~1.}
         \label{FigGam}
    \end{figure}
   \begin{table*}
      \caption[]{Field centers, observation intervals and exposure times of 
                 pointed observations on SMC}
            \begin{flushleft}
            \begin{tabular}{lccrrr}
            \hline
            \hline
            \noalign{\smallskip}
                    &                     &                     &                        &                 &                 \\
              Field & R.A.(2000)          & Decl.(2000)         & time start [UT]        & time end [UT]   & exposure [ksec] \\
            \noalign{\smallskip}
            \hline
            \noalign{\smallskip}
      A1    & $\rm 0^h58^m12.0^s$ & $\rm -72^d16^m48^s$ & ~8-Oct-91 -- 03:10  &  9-Oct-91 -- 02:47 & 17.0   \\
      A2    & $\rm 0^h58^m12.0^s$ & $\rm -72^d16^m48^s$ & 17-Apr-92 -- 17:07  & 27-Apr-92 -- 16:34 & ~9.7   \\
      B1    & $\rm 0^h50^m45.5^s$ & $\rm -73^d13^m48^s$ & ~9-Oct-91 -- 03:03  &  9-Oct-91 -- 04:43 & ~1.4   \\
      B2    & $\rm 0^h50^m45.5^s$ & $\rm -73^d13^m48^s$ & 15-Apr-92 -- 15:40  & 24-Apr-92 -- 18:17 & 22.7   \\
      C     & $\rm 1^h13^m24.0^s$ & $\rm -72^d49^m12^s$ & 16-Oct-91 -- 07:33  & 19-Oct-91 -- 23:32 & 22.1   \\
      D     & $\rm 1^h05^m55.2^s$ & $\rm -72^d33^m36^s$ & 10-Apr-93 -- 11:54  & 14-Oct-93 -- 16:54 & 30.9   \\
      E     & $\rm 0^h54^m28.7^s$ & $\rm -72^d45^m36^s$ & ~9-May-93 -- 07:17  & 12-May-93 -- 20:14 & 16.8   \\
      F1    & $\rm 0^h42^m55.2^s$ & $\rm -73^d38^m24^s$ & ~5-Dec-92 -- 23:58  & 8-Dec-92  -- 01:40 & 9.7    \\
      F2    & $\rm 0^h42^m55.2^s$ & $\rm -73^d38^m24^s$ & ~5-Apr-93 -- 06:12  & 26-Apr-93 -- 23:31 & ~8.3   \\
      G1    & $\rm 1^h01^m16.7^s$ & $\rm -71^d49^m12^s$ & ~6-Dec-92 -- 09:33  & 6-Dec-92 -- 13:10 & ~3.6    \\
      G2    & $\rm 1^h01^m16.7^s$ & $\rm -71^d49^m12^s$ & ~7-Oct-93 -- 20:06  &
 10-Oct-93 -- 14:46 & ~4.6   \\
     X1     & $\rm 0^h37^m19.2^s$ & $\rm -72^d14^m24^s$ & 28-Apr-92 -- 12:50  & ~3-Jun-92 -- 17:58 & 6.4    \\
     X2     & $\rm 0^h37^m19.2^s$ & $\rm -72^d14^m24^s$ & ~2-Oct-93 -- 10:15 & ~9-Oct-93 -- 06:06 & 1.7     \\
     Y1     & $\rm 0^h58^m33.5^s$ & $\rm -71^d36^m00^s$ & 29-Mar-93 -- 08:07  & 30-Mar-93 -- 21:10 & 5.2    \\
     Y2     & $\rm 0^h58^m33.5^s$  & $\rm -71^d36^m00^s$ & ~1-Oct-93 -- 14:06 & ~9-Oct-93 -- 19:19 & 7.2    \\ 
            \noalign{\smallskip}
           \hline
           \end{tabular}
           \end{flushleft}
   \end{table*}

\subsection{Detections and Count Rates}

A sophisticated procedure and technique has been used to generate a catalog of
X-ray detections from the 9 pointings listed in Table~1. This procedure is 
described in detail in a separate paper. A maximum likelihood detection 
algorithm in {\it EXSAS} (Zimmermann et al. 1993) has been used to determine 
the position of the source and the counts in the standard \ros energy bands 
Soft=(channel 11-41, roughly 0.1-0.4~keV), Hard=(channel 52-201, roughly 
0.5-2.1~keV), Hard1=(channel 52-90, roughly 0.5-0.9~keV), Hard2=(channel 
91-201, roughly 0.9-2.0~keV). Hardness ratios HR1 and HR2 have been calculated 
from the counts in the bands HR1=(H-S)/(H+S) and HR2=(H2-H1)/(H1+H2). The 
result is given in Table~3 and Table~4. Upper limit counts have been 
calculated by a maximum likelihood algorithm (cf. Pollock et al. 1981). The 
time dependent X-ray count rate of the X-ray binaries in the \ros band 
(0.1-2.4~keV) has been studied for time scales in excess of individual 
observation intervals (typically 10 minutes to 1 hour). Variability on shorter
time scales has not been looked for with the exception of SMC~X-1 where two 
X-ray bursts (or flares) have been discovered and studied. Most sources have 
been covered by more than one pointing and a search for long-term ($\sim$half 
a year) variability became possible.

\subsection{Selection Criteria for Hard Sources}

The selection criteria for hard X-ray binaries were set such that, first of
all they must comprise all known hard X-ray binaries in the SMC and second 
they must cover all sources with an X-ray luminosity above $\sim$3$\rm \times 
10^{35} erg\ s^{-1}$ 
(0.1 - 2.4 keV) and detectable temporal variability. The X-ray luminosity has 
been determined from the X-ray counts applying a conversion factor 

$$Luminosity[erg/s]\ =\ count rate [cts/s]\times1.67\times10^{37},$$  

appropriate for a thermal bremsstrahlung spectrum of temperature 2~keV, a 
galactic absorbing column density of $\rm 3\times 10^{20} cm^{-2}$ and a 
SMC intrinsic absorption column of $\rm 4\times 10^{21} cm^{-2}$ (cf.
absorbing columns given in Figure~7 and the smaller X-ray absorption in the
SMC per H-atom discussed below) and assuming a distance to the SMC of 65~kpc 
(cf. Wang \& Wu 1992). As many of the SMC X-ray binaries may be (and indeed 
turn out to be) heavily absorbed (cf. Table~6), a rather large SMC intrinsic 
absorption of $\rm 4\times 10^{21}\ cm^{-2}$ seems to be justified. But it 
introduces some uncertainty in the luminosity threshold due to the unknown 
intrinsic luminosity. On the other hand a simple conversion from X-ray count 
rates to luminosities and assuming an a priori absorption (like the galactic 
foreground value) would just give a lower limit on the luminosity.

\begin{table}
  \caption[]{Selection criteria for spectrally hard and soft X-ray binary 
             candidates}
  \begin{flushleft}
  \begin{tabular}{ccc}
  \hline
  \hline
  \noalign{\smallskip}
  Quantity          & \multicolumn{2}{c}{Selection criterion}       \\
                    & hard source           & soft source           \\
  \noalign{\smallskip}
  \hline
  \noalign{\smallskip}
    HR1             & $\rm \ge 0.5$         & $\rm \le -0.8$        \\
    HR2             & $\rm \ge 0.0$         & --                    \\
  count rate        & $\rm > 0.015\ s^{-1}$ & $\rm > 0.015\ s^{-1}$ \\
  extent likelihood & $\rm < 50$            & --                    \\
  time variability  & $\rm >\ hours$        & --                    \\
  \noalign{\smallskip}
  \hline
  \end{tabular}
  \end{flushleft}
\end{table}

We considered all sources fulfilling the selection criteria listed in 
Table~2. These sources have a value for the hardness ratio $\rm HR1\ge 
+0.5$ (excluding spectrally softer sources like stars) a count rate above 
$\rm 0.015\ counts\ s^{-1}$ and a likelihood of extension $\rm <$50 
(determined by a standard maximum likelihood detection, Zimmermann et al. 
1993). The extent selection has been applied in order to discriminate 
supernova remnants or other extended structure from point like sources. 
Another criterion ($\rm HR2<0$) has been introduced in order to discriminate 
against spectrally softer sources (e.g. AGNs). Errors in the hardness ratios 
HR1 and HR2 have not been considered as they turn out to be small for the 
considered count rates. Two sources have been rejected as they coincide 
with struts of the detector system. 13 sources fulfilled these criteria 
(cf. Table~3 and Figure~1). Of the seven sources known from \ein observations 
(with the catalog numbers 60, 83, 84, 103, 105, 153, and 242, cf. Table~3) 
only SMC~X-1 (number 242) and 1~E0050.1-7248 (RX~J0051.8-7231, number 83) 
have been considered by Wang \& Wu (1992) as X-ray binaries. Sources with 
the indices 103, 105, and 153 will be discussed in section~4.4. Six sources 
(with catalog numbers 3, 19, 26, 69, 100, and 158) have not been seen with 
\ein . One of them is SMC~X-2, a Be type transient, the other five are \ros 
discoveries and good candidate X-ray binaries. 

Finally accepted sources had to show variability in time detectable either 
within one observation with time scales of minutes to days or from one 
pointing to another with a time scale of about half a year. The latter 
time scale allowed us to discover two X-ray transients not seen with \ein . 
All new hard X-ray binary candidates still lack an optical identification, 
but optical identification work is in progress (cf. sections 3.4.3 and 3.4.5). 
Optical identification will clarify that we did not confuse them with variable
background AGNs. The seven accepted sources (showing time variability) and 
the six rejected sources (due to missing time variability) are listed in 
Table~3 and are marked in Figure~1.

\subsection{Candidate supersoft X-ray sources}

To search for candidate supersoft sources we selected sources with $\rm HR1 
< -0.8$ and a count rate $\rm \ge 0.015\ counts\  s^{-1}$. We found five 
candidates. Four of these were presented by Kahabka et al. (1994). The new 
candidate is RX~J0103.8-7254. A correlation with the star CPD-73~2349 as 
stated by Wang \& Wu (1992) is in error. There are two ot three not too faint 
blue stars in the error circle (M.Pakull, private communication). The X-ray 
source has also been detected by the \ros {\sl WFC} and is contained in the 
{\it 2RE} source catalog (Pye et al. 1995). This would be in favor for a 
galactic nature of the object with a comparatively low foreground absorbing 
column density. We searched in our SMC point source catalog (Kahabka \& 
Pietsch 1996) for CAL~87 like candidates, which appear to be hotter than the 
ordinary SSS and have HR1$>$0 and HR2$<$-0.6 and found none. 

\section{Observational results of individual sources}

We report about the results obtained from the {\sl RASS} and the pointed \ros 
observations. The {\it RASS} observations are typically a factor of 10 less 
deep in source flux and besides SMC~X-1 no SMC X-ray binary was detected. We 
can only give upper limit count rates for the sample of binaries found from 
previous observations or during the deep pointed observations. 

The result of the selection defined in the previous chapter and applied to 
the pointed data is given in Table~3 to Table~5. We reject six hard X-ray 
binary candidates (due to amissing time variability) and select seven 
candidates (see Table~3). From the sample of supersoft candidates we reject 
1~source and select 4~sources (see Table~4). Finally we give in Table~5 upper 
limits to count rates for those pointings where the discovered transient hard 
X-ray sources have not been detected and give upper limits to count rates for 
two previously recorded transient hard X-rays sources contained in the field 
of view of our observations. The result of detailed spectral fitting of the 
brighter hard X-ray sources is given in Table~6. For the supersoft sources the
result of the spectral modeling has already been published in Kahabka et al. 
(1994). A compilation (catalog) of the presently known and published sample of
SMC X-ray binaries and candidates is given in Table~10.

   \begin{table*}
      \caption[]{Equatorial coordinates, error radius, count rate 
                 (0.1-2.4 keV), hardness ratio HR1 and HR2, off-axis angle 
                 for the 7 accepted (time variable) and the 6 rejected 
                 (showing no time variability) candidate hard X-ray binaries
                 in the SMC. The parameters are deduced for each source 
                 separately and may differ (slightly) from the parameters  
                 in the SMC point source catalog (Kahabka \& Pietsch 1996).}
            \begin{flushleft}
            \begin{tabular}{llllccccc}
            \hline
            \hline
            \noalign{\smallskip}
  catalog &              &                  &           & Err.rad.$\rm ^b$  
 & count rate     & hardness ratio & off-axis &    \\
  number  & Source name  & R.A.(2000)       & Decl.(2000)         & [arcsec]  
& $\rm [sec^{-1}]$ &  HR1/HR2      & [arcmin] & Remarks$\rm ^a$ \\
            \noalign{\smallskip}
            \hline
            \noalign{\smallskip}
 \multicolumn{9}{c} {accepted X-ray binaries (variable in time)} \\
            \noalign{\smallskip}
            \hline
            \noalign{\smallskip}
  242 & SMC X-1       & $\rm 1^h17^m02.2^s$ & $\rm -73^d26^m33^s$ & 65     
& 0.373$\pm$0.003       & +1 /           & 41  & 63 [2]       \\
      &               &                     &                     &          
&                 & 0.317$\pm$0.007   &     &                 \\
  100 & SMC X-2       & $\rm 0^h54^m31.3^s$ & $\rm -73^d40^m54^s$ & 62     
& 0.390$\pm$0.022       & 0.971$\pm$0.028 / & 32  & --        \\
      &               &                     &                     &          
&                 & 0.637$\pm$0.043   &     &                 \\
    3 & RX J0032.9-7348 & $\rm 0^h32^m55.1^s$ & $\rm -73^d48^m11^s$ & 62   
& 0.122$\pm$0.005        & +1/            & 43  &   --        \\
      &               &                     &                     &          
&                  & 0.546$\pm$0.038  &     &                 \\         
   69 & RX J0049.1-7250 & $\rm 0^h49^m04.6^s$ & $\rm -72^d50^m53^s$ & 22     
& 0.042$\pm$0.008        & +1/            & 24  &   --        \\
      &               &                     &                     &          
&                 & 0.836$\pm$0.100   &     &                 \\
   83 & RX J0051.8-7231 & $\rm 0^h51^m53.0^s$ & $\rm -72^d31^m45^s$ & 11     
& 0.112$\pm$0.003        & 0.951$\pm$0.011 / & 18  & 27 [2] \\
      &               &                     &                     &          
&                 & 0.479(0.023)   &     &                    \\
   84 & RX J0052.1-7319 & $\rm 0^h52^m11.3^s$ & $\rm -73^d19^m13^s$ & 11     
& 0.0214$\pm$0.011      & 0.941$\pm$0.015/  &  8  & 29 [2]  \\
      &               &                     &                     &          
&                 & 0.612$\pm$0.014   &     &                 \\
  158 & RX J0101.0-7206 & $\rm 1^h01^m03.2^s$ & $\rm -72^d06^m57^s$ & 5      
& 0.048$\pm$0.002        & 0.882$\pm$0.036/  & 16  & --     \\
      &               &                     &                     &          
&                 & 0.472$\pm$0.037   &     &                 \\
            \noalign{\smallskip}
            \hline
            \noalign{\smallskip}
\multicolumn{9}{c} {rejected X-ray binaries (missing time variability)} \\
            \noalign{\smallskip}
            \hline
            \noalign{\smallskip}
  19 & RX~J0036.9-7138 & $\rm 0^h36^m59.9^s$ & $\rm -71^d38^m07^s$ & 68 & 0.0246$\pm$0.0029 & +1              & 36 & -- \\
     &                 &                     &                     &    &                   & 0.383$\pm$0.116 &    &    \\
  26 & RX~J0038.6-7310 & $\rm 0^h38^m36.3^s$ & $\rm -73^d10^m22^s$ & 65 & 0.0317$\pm$0.0025 & +1/             & 34 & -- \\
     &                 &                     &                     &    &                   & 0.393$\pm$0.075 &    &    \\
  60 & RX~J0047.5-7308 & $\rm 0^h47^m30.8^s$ & $\rm -73^d08^m46^s$ & 63 & 0.0351$\pm$0.0021 & +1/             & 38 & [1], 16 [2] \\
     &                 &                     &                     &    &                   & 0.209$\pm$0.057 &    &    \\               
 103 & RX~J0054.9-7226 & $\rm 0^h54^m57.3^s$ & $\rm -72^d26^m39^s$ & 11 & 0.0273$\pm$0.0012 & +1/             & 18 & 35 [2] \\
     &                 &                     &                     &    &                   & 0.554$\pm$0.036 &    &    \\
 105 & RX~J0055.4-7210 & $\rm 0^h55^m29.2^s$ & $\rm -72^d10^m53^s$ & 11 & 0.0247$\pm$0.0010 & +1/             & 14 & 36 [2] \\
     &                 &                     &                     &    &        & 0.359$\pm$0.040    &    &        \\
 153 & RX~J0100.7-7211 & $\rm 1^h00^m43.4^s$ & $\rm -72^d11^m34^s$ & 11 & 0.0240$\pm$0.0011 & 0.969           & 13 & 45 [2] \\
     &                 &                     &                     &    &        & 0.329$\pm$0.041    &             \\ 
           \noalign{\smallskip}
           \hline
           \noalign{\smallskip}
           \end{tabular}
           \end{flushleft}
a) [1] SNR~0045-734. [2] Source number from catalog of discrete X-ray 
sources of Wang \& Wu (1992). \\
b) For sources with off-axis angles in excess of $\rm \sim 30\amin$ a 
systematic error in position of $\rm 1 arcmin$ has been added quadratically 
to the positional error determined with the maximum likelihood algorithm.
   \end{table*}

\begin{table*}
  \caption[]{Equatorial coordinates, error radius, count rate (0.1-2.4 keV),
             hardness ratio HR1 and HR2, off-axis angle for the 4 selected
             and 1 candidate supersoft source.}
  \begin{flushleft}
  \begin{tabular}{rcccccccc}
  \hline
  \hline
  \noalign{\smallskip}
catalog &             &            &                     & Err.rad. & count rate   & hardness ratio & off-axis & Remarks$\rm ^a$ \\
number  & Source name & R.A.(2000) & Decl.(2000)         & [arcsec] & [sec$\rm ^{-1}$] & HR1/HR2         & [arcmin] &         \\
  \noalign{\smallskip}
  \hline
  \noalign{\smallskip}
\multicolumn{9}{c} {selected~supersoft~sources} \\
  \noalign{\smallskip}
  \hline
  \noalign{\smallskip}
 20 & 1E~0035.4-7230  & $\rm 0^h37^m19.5^s$ & $\rm -72^d14^m08^s$ & 11       & 0.504$\pm$0.010    & -0.966/-0.954  & 0.2      & variable        \\
135 & 1E~0056.8-7146   & $\rm 0^h58^m37.2^s$  & $\rm -71^d35^m56^s$ & 11     & 0.356$\pm$0.007     & -0.993/-1      & 0.3      & PN        \\
 62 & RX J0048.4-7332 & $\rm 0^h48^m16.3^s$ & $\rm -73^d31^m45^s$ & 11      & 0.188$\pm$0.003     & -0.973/-0.923  & 21       & symb. nova        \\
133 & RX J0058.6-7146 & $\rm 0^h58^m35.8^s$ & $\rm -71^d46^m02^s$ & 13      & 0.0256$\pm$0.0025   & -0.994/-1      & 10       & transient        \\
  \noalign{\smallskip}
  \hline
  \noalign{\smallskip}
\multicolumn{9}{c} {candidate~supersoft~sources} \\
  \noalign{\smallskip}
  \hline
  \noalign{\smallskip}
181 & RX~J0103.8-7254 & $\rm 1^h03^m53.4^s$ & $\rm -72^d54^m49^s$ & 15 & 0.0218 & -1/0.0 & 23 & [1] \\
  \noalign{\smallskip}
  \hline
  \noalign{\smallskip}
  \end{tabular}
  \end{flushleft}
a) source [1] correlates with the {\it EXOSAT} source EXO~0102.3-7318 (cf. 
Table~2B from Wang \& Wu 1992). The source is not yet optically identified.  
\end{table*}

\subsection{Temporal Analysis}

\begin{table*}
  \caption[]{\ros pointed $\rm 2\sigma$ upper limit count rates (0.1-2.4~keV)
             and (0.5-2.0~keV) for the detected and the recorded transient 
             X-ray sources. The count rate for a detection is given for 
             RXJ0049.1-7250 in the (0.5-2.0~keV) band.}
  \begin{flushleft}
  \begin{tabular}{lccrrrc}
  \hline
  \hline
  \noalign{\smallskip}
    Source name & \multicolumn{2}{c} {$\rm 2\sigma$ upper limit}       & time start & time end & offax    & field \\
                & \multicolumn{2}{c} {count rate [$\rm 10^{-3}\ counts\ s^{-1}$]} & 
 &          & [arcmin] &       \\
                & (0.1-2.4~keV)       &       (0.5-2.0~keV)        &            &          &          &       \\
  \noalign{\smallskip}
  \hline
  \noalign{\smallskip}
\multicolumn{7}{c} {detected~transient~X-ray~sources} \\ 
  \noalign{\smallskip}
  \hline
  \noalign{\smallskip}
    SMC X-2      & 0.58 & 0.66 & 15-Apr-92 -- 15:40 & 24-Apr-92 -- 18:17 
& 31.5 & B2 \\
  RX~J0049.1-7250 & 2.85 & 3.50($\rm \pm 0.59$) & 15-Apr-92 -- 15:40 & 24-Apr-92 -- 18:17
& 24.0 & B2 \\
  RX~J0051.8-7231 & 2.70 & 2.69 &  8-Oct-91 -- 03:03 &  9-Oct-91 -- 04:47      
& 32.3 & A1 \\
                 & 3.16 & 3.04 & 17-Apr-92 -- 17:07 & 27-Apr-92 -- 16:34      
& 32.3 & A2 \\
  RX~J0052.1-7319 & 2.21 & 1.59 &  9-Oct-91 -- 03:03 &  9-Oct-91 -- 04:43     
&  8.2 & B1 \\
  RX~J0101.0-7206 & 1.46 & 0.62 & 17-Apr-92 -- 17:07 & 27-Apr-92 -- 16:34
& 16.4 & A2 \\
                & 1.44 & 1.37 &  6-Dec-92 -- 09:33 &  6-Dec-92 -- 13:10
& 17.8 & G1 \\
                & 2.58 & 2.76 &  7-Oct-93 -- 20:06 & 10-Oct-93 -- 14:45
& 17.8 & G2 \\ 
  \noalign{\smallskip}
  \hline
  \noalign{\smallskip}
\multicolumn{7}{c} {recorded~but~not~detected~transient~X-ray~sources} \\ 
  \noalign{\smallskip}
  \hline
  \noalign{\smallskip}
SMC X-3      & 1.73 & 1.17 &  8-Oct-91 -- 03:10 &  9-Oct-91 -- 02:47  
& 29.0 & A1\\
             & 0.59 & 0.42 & 17-Apr-92 -- 17:07 & 27-Apr-92 -- 16:34  
& 29.0 & A2\\
             & 1.48 & 0.55 &  9-May-93 -- 07:26 & 12-May-93 -- 20:10  
& 22.5 & E\\
RX J0059.2-7138 & - & 0.54 &  8-Oct-91 -- 03:10 &  9-Oct-91 -- 02:47  
& 38.2 & A1\\
             &    - & 1.23 & 17-Apr-92 -- 17:07 & 27-Apr-92 -- 16:34  
& 38.2 & A2\\
             & 2.04 & 0.71 &  6-Dec-92 -- 09:33 &  6-Dec-92 -- 13:10 
& 14.2 & G1\\
             & 1.05 & 0.48 &  7-Oct-93 -- 20:06 & 10-Oct-93 -- 14:46 
& 14.2 & G2\\
             & 1.04 & 0.51 & 29-Mar-93 -- 08:07 & 30-Mar-93 -- 21:03  
&  4.2 & Y1\\
             & 0.91 & 0.25 &  1-Oct-93 -- 14:06 &  9-Oct-93 -- 19:19 
&  4.2 & Y2\\
  \noalign{\smallskip}
  \hline
  \end{tabular}
  \end{flushleft}
\end{table*}

The two recorded transients SMC~X-3 and the 2.8~s transient RX~J0059.2-7138
recently discovered by Hughes (1994) have not been detected in our pointings. 
The upper limit luminosities of $\rm \sim9\times10^{34}\ erg\ s^{-1}$ and
$\rm \sim6\times10^{34}\ erg\ s^{-1}$ respectively (in case of a SMC~X-2 like 
spectrum, cf. Table~5) are consistent with a Be type transient nature of
these two sources.    

Light curves have been generated for the five brightest transient and 
persistent hard X-ray binaries in the energy interval 0.1-2.4~keV. One data 
point has been drawn per observation interval (of duration $\sim$10~min to 
1~hour).

   \begin{figure*}
      \centering{
      \vbox{\psfig{figure=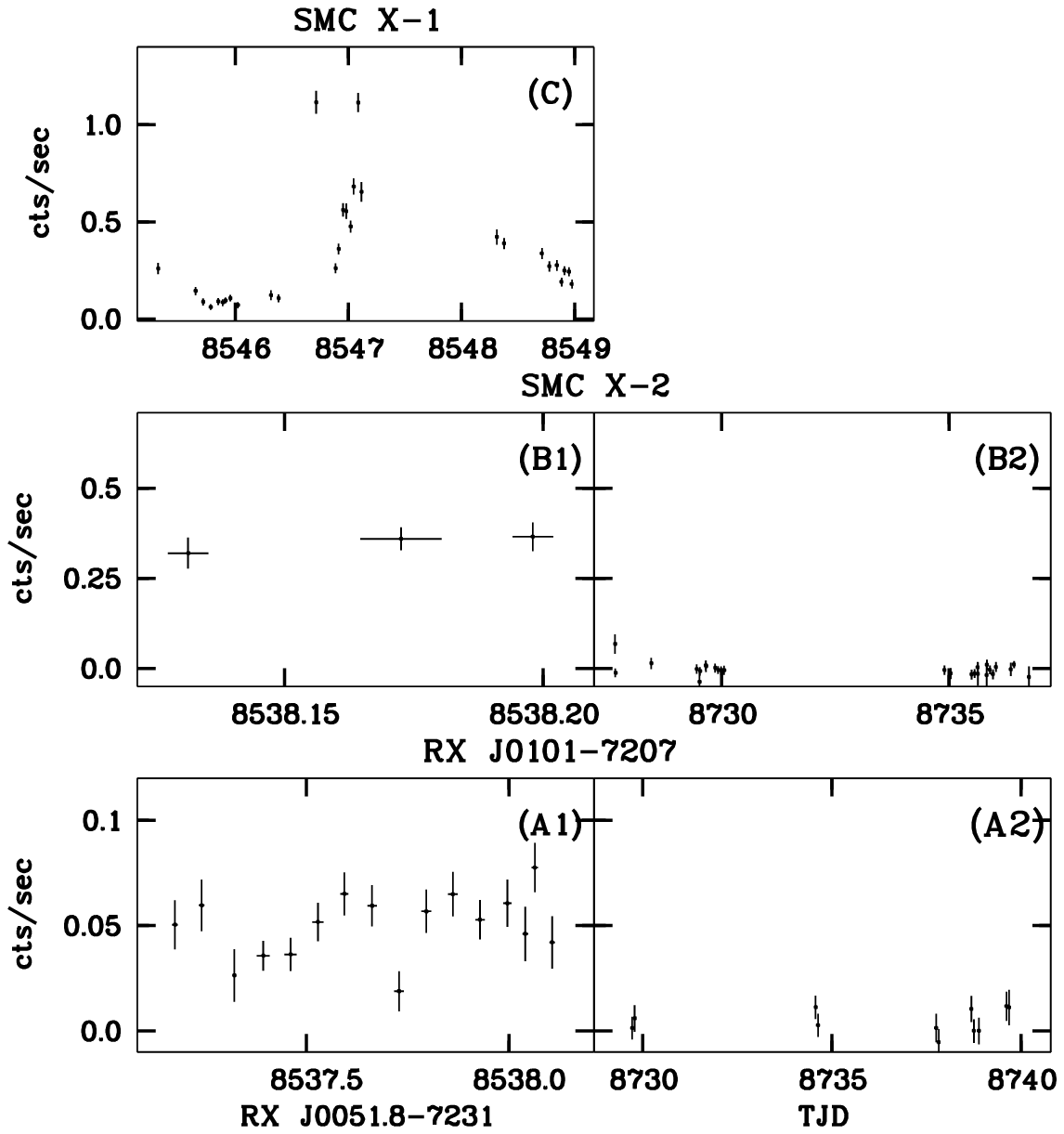,width=17.5cm,%
       bbllx=3.0cm,bblly=1.0cm,bburx=17.0cm,bbury=14.0cm,clip=}}\par
       }
      \centering{
      \vbox{\psfig{figure=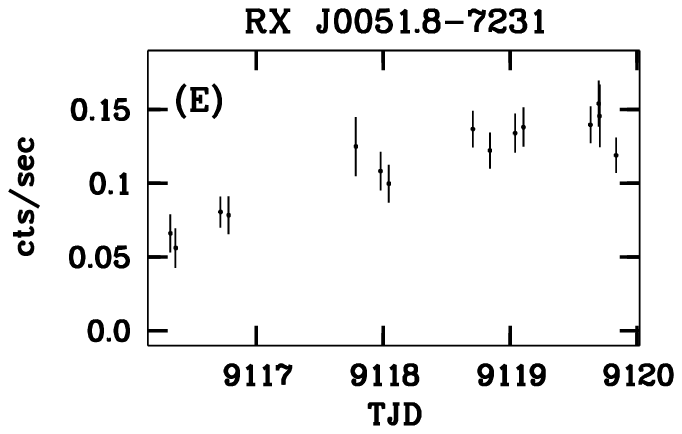,width=17.5cm,%
       bbllx=3.cm,bblly=8.5cm,bburx=17.0cm,bbury=13.cm,clip=}}\par
       }
      \caption[]{Light curves (0.1 - 2.4 keV) of SMC~X-1 (pointing C), SMC~X-2 
                 (pointings B1 and B2), RX~J0101.0-7206 (pointings A1 and A2) 
                 and RX~J0051.8-7231 (pointing E). For each observation 
                 interval one data point is drawn. Time TJD is given 
                 (JD - 2440000.5).}
         \label{FigGam}
    \end{figure*}

SMC~X-1 has been observed during one full binary orbit (cf. Figure~3 and 
Figure~8). The observation took place during a low intensity state. It starts 
with a mean count rate of $\rm 0.16\ counts \ s^{-1}$ covering an extended 
X-ray eclipse. The source then showed a flare like event with an increase in 
count rate by a factor of 10 up to $\rm 1.5\ counts \ s^{-1}$ in the peak (cf.
Figure~8). The full rise of the peak has not been observed but an increase by 
a factor of 2 in intensity occurred within 200~seconds. The source then showed
a flare with a rise time of about 6~hours and a second intensity increase (at 
orbital phase $\sim$0.34) with a similar peak count rate as in the first 
flare. The separation of the two intensity spikes is $\rm \sim 9.6~hours$. 
The second peak has not been seen during the rise time but during a plateau 
(of $\rm \sim 250~sec$ duration). In the last part of the observation the 
intensity declines from $\rm \sim0.5 \ counts\ s^{-1}$ to the initial observed
level of $\rm \sim0.2\ counts\ s^{-1}$.

The X-ray transient SMC~X-2 has been discovered with \ros in its second 
outburst after {\sl SAS-3} (Clark et al. 1978) in pointing B1 for 1.8~hours 
and with a mean count rate of 0.39\ counts\ ~$\rm s^{-1}$ (cf. Figure~3). Half
a year after this \ros observation the X-ray source was no longer detected. A 
$\rm 2\sigma$ upper limit to the count rate of $\rm 5.8\times 10^{-4}\ counts\ 
s^{-1}$ has been deduced indicating variability by a factor $\rm >6\times 
10^3$.

The source RX~J0051.8-7231 has already been discovered in \ein observations 
(\ein name 1E~0050.1-7248, Wang \& Wu 1992). It was in the field of view of
the \ros pointings A1, A2 and E but has only been detected in pointing E 
which indicates intensity variations by at least a factor of 35 (cf. Table~3 
and Table~5). In pointing E an increase in intensity by a factor of three has
been observed within three days (cf. Figure~3).

RX~J0052.1-7319 has been discovered in \ein observations (\ein name 
1E~0050.3-7335, Wang \& Wu 1992). During \ros observations the source has been
found to be variable (cf. Figure~4). From \ein observations no variability 
has been reported. The source has been detected in pointing B2 with a mean 
count rate of $\rm 0.089 \pm 0.02\ counts\ s^{-1}$.

\begin{figure*}
  \centering{
  \vbox{\psfig{figure=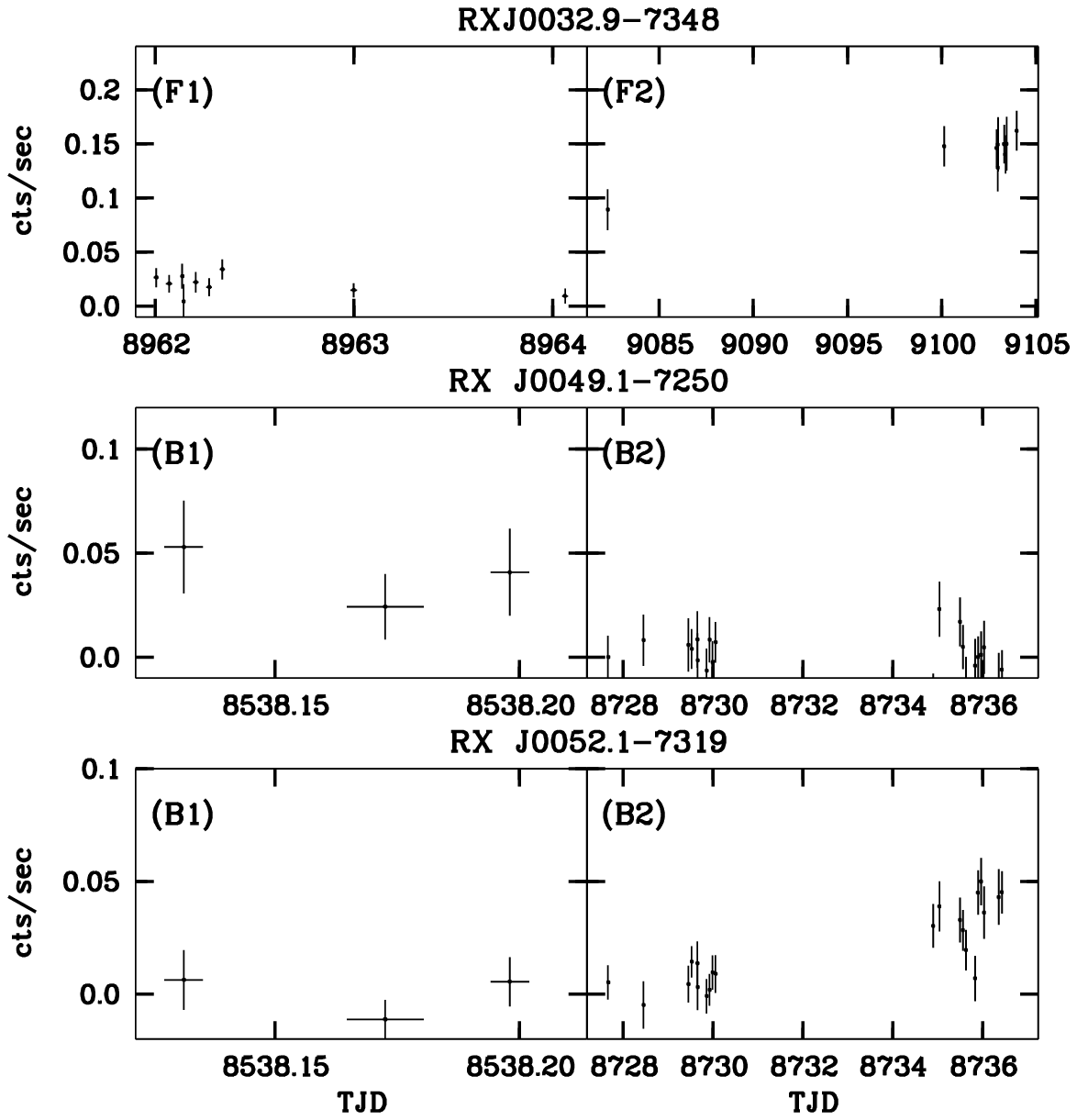,width=17.5cm,%
        bbllx=3.0cm,bblly=0.5cm,bburx=17.0cm,bbury=14.0cm,clip=}}\par
            }
  \caption[]{Light curves (0.5-2.4~keV) of RX~J0032.9-7348 (pointings F1 and
             F2) and light curves (0.1-2.4~keV) of RX~J0049.1-7250 (pointings 
             B1 and B2) and RX~J0052.1-7319 (pointings B1 and B2). For each 
             observation interval one data point is drawn. Time is given in 
             TJD (JD - 2440000.5).}
\end{figure*}

A new X-ray transient (RX~J0101.0-7206) has been discovered with \ros at the 
north-eastern boundary of the HII region N66. The transient has been seen in 
outburst in pointing A1 for 22 hours (cf. Figure~3). Half a year later the 
X-ray source has no longer been detected. A $\rm 2\sigma$ upper limit count 
rate of $\rm 1.46\times 10^{-3}\ counts\ s^{-1}$ was deduced indicating a 
variability by a factor of $\rm >33$ (cf. Table~3 and Table~5).

Another highly variable X-ray source RX~J0049.1-7250 was discovered with \ros 
north west of the supernova remnant N~19. The variability from pointing B1 to 
B2 was a factor of 12 in the (0.5-2.0~keV) band (cf. Figure~4). The source is 
highly absorbed (cf. Table~6) and has not been detected in the soft band.

RX~J0032.9-7348 is a candidate for a variable X-ray binary at the south-eastern
border of the SMC. It showed a count rate of $\rm 0.122\ counts\ s^{-1}$ in 
pointing F2. In another pointing taken $\sim$0.4~years before the source was 
about a factor of 6 weaker.

\subsection{Spectral Analysis}

A thermal bremsstrahlung spectral fit has been applied to the X-ray spectra. 
For the absorption (in our Galaxy and the SMC) two different abundances have 
been used. A spectral fit has been performed first assuming cosmic abundances 
and second cosmic abundances for a galactic contribution of 
$\rm 3\times10^{20}\ cm^{-2}$ and reduced metallicities (by a factor of about 
7, cf. Pagel 1993) for the additional SMC contribution). 

    \begin{figure}  
      \centering{
      \vbox{\psfig{figure=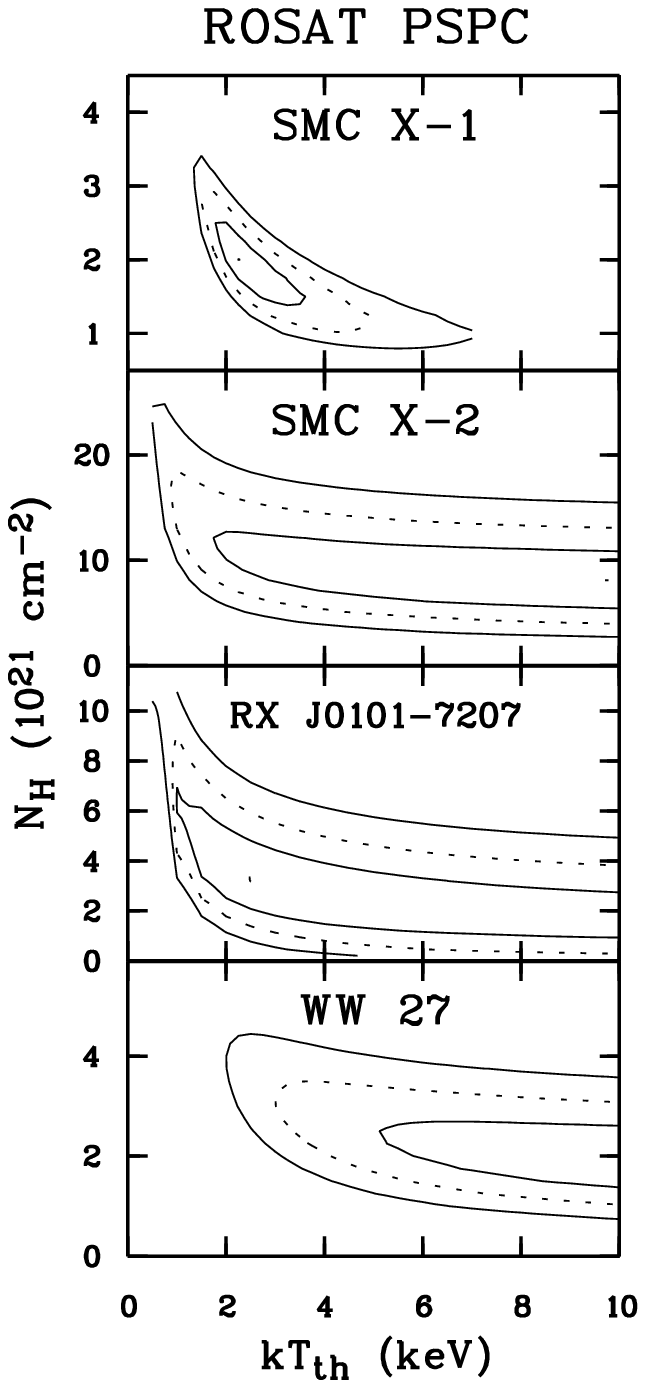,width=9cm,%
       bbllx=9.5cm,bblly=4.5cm,bburx=16.5cm,bbury=19cm,clip=}}\par
                }
      \caption[]{68, 95, and 99\% confidence parameter plane for the hydrogen 
                 absorbing column density within the SMC (assuming SMC metal 
                 abundances reduced by a factor of $\sim$7 compared to 
                 galactic opacities) with a galactic absorption term of 
                 $\rm 3\times 10^{20} cm^{-2}$ and the temperature of 
                 a thermal bremsstrahlung fit applied to the spectra of the
                 four X-ray binaries SMC~X-1, SMCX~2, RX~J0101.0-7206 and 
                 RX~J0051.8-7231 (WW27) (from top to bottom).}
         \label{FigGam}
    \end{figure}
    \begin{figure}
      \centering{
      \vbox{\psfig{figure=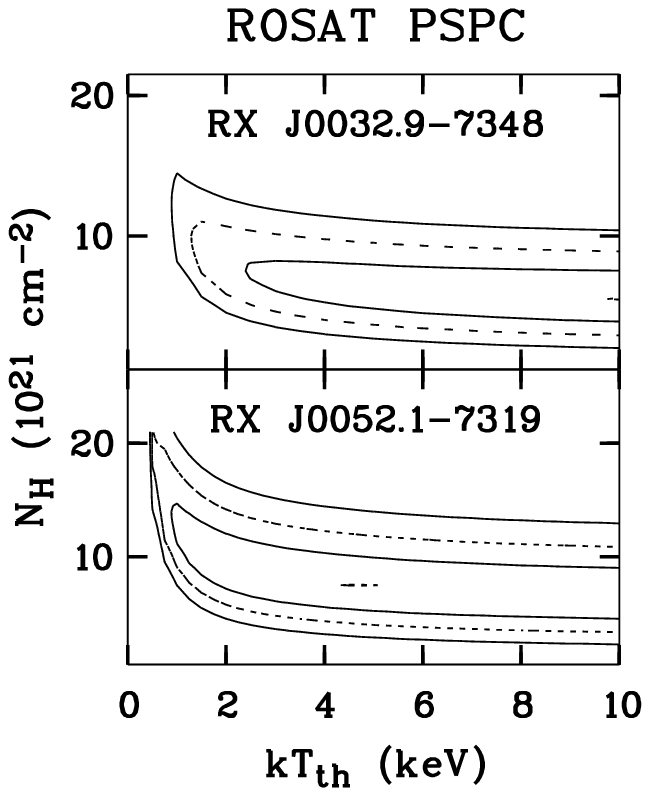,width=9cm,%
      bbllx=9.5cm,bblly=10.5cm,bburx=17.0cm,bbury=19.2cm,clip=}}\par
                }
      \caption[]{As in Fig.5, but for the two candidate X-ray binaries
                 RX~J0032.9-7348 and RX~J0052.1-7319.}
    \end{figure}

The model of reduced metallicities in the absorbing opacities gives higher
hydrogen column densities (by $\sim$70\%), somewhat higher bremsstrahlung
temperatures (of $\sim$25\%) and lower X-ray luminosities (by $\sim$25\%).
The result of the spectral fitting is detailed in Table~6. We give the 
exposure time, the best fit thermal bremsstrahlung temperature (with 95\%
confidence errors), the absorbing hydrogen column density (with 95\% 
confidence errors), the unabsorbed and absorbed X-ray luminosity 
(0.15-2.4~keV), the chi-squared and the degrees of freedom for the fit.
In Figure~5 and Figure~6 the 68, 95, 99 $\%$ confidence parameter contours
for low metal abundance absorption within the SMC versus thermal bremsstahlung
temperature for six SMC hard X-ray binaries is shown. With the exception of 
SMC~X-1 the \ros {\sl PSPC} can limit the thermal bremsstrahlung temperature 
only at the lower bound. In our low-state observation SMC~X-1 has a 
temperature below $\sim$5~keV (95\% confidence). The best-fit temperature 
is for all sources above 1~keV and for RX~J0051.8-7231 (WW~27) above 10~keV. 
Temperatures above $\sim$10~keV are typical for HMXBs and temperatures below 
$\sim$10~keV for LMXBs. This may be due to reprocessed X-rays determining the 
spectra of the latter class. One has to note that the spectrum of SMC~X-1 in 
our low-state observation may be dominated by reprocessed radiation. The high 
absorbing hydrogen column deduced from nearly all X-ray spectra can be 
explained by matter located in the SMC and in part within the individual 
binary system. In Figure~7 the positions of the discussed X-ray binaries are 
drawn on a HI column density map deduced from 21-cm data (Luks 1994). This 
illustrates the largest SMC column expected to be deduced from the X-ray 
spectra of these sources.

In Table~7 the hydrogen column densities of the X-ray sources as deduced 
from a spectral fit are given together with the values found from the HI 
map (smoothed with a Gaussian $\rm \sigma$ of 8'). One has to admit that 
the resolution of the radio map with $\rm \sim 15\amin$ is rather coarse. 
We also refer to the work of Bessell (1991) (and references given therein) 
for a discussion of the SMC foreground and the SMC intrinsic reddening.

   \begin{table*}
      \caption[]{Spectral parameters of a thermal bremsstrahlung fit for 
                 the 7 X-ray binaries in the SMC. Thermal bremsstrahlung
                 temperature $\rm kT_{th}$ (keV), galactic absorbing column 
                 density ($\rm 10^{21} cm^{-2}$), X-ray luminosity
                 $\rm L_{x}$ (0.15 - 2.4 keV, $\rm 10^{37} erg/s$, 65 kpc), 
                 absorbed X-ray luminosity $\rm L_{abs}$ (0.15 - 2.4 keV, 
                 $\rm 10^{37} erg/s$, 65 kpc, excluding a galactic 
                 contribution of $\rm 3.\times 10^{20} cm^{-2}$), $\rm\chi_2$ 
                 and degrees of freedom (DOF) of fit, and pointing.}
            \begin{flushleft}
            \begin{tabular}{lccccccc}
            \hline
            \hline
            \noalign{\smallskip}
            \multispan{8}
             galactic~opacities \\
            \noalign{\smallskip}
            \hline
            \noalign{\smallskip}
              & $\rm kT_{th}$  & $\rm N_H$               & $\rm L_{x}$                & $\rm L_{abs}$     &      &      \\
 Source       & [keV]      & [$\rm 10^{21} cm^{-2}$] & [$\rm 10^{37} erg/s$]  & [$\rm 10^{37} erg/s$] & $\rm\chi_2$ & DOF & Field \\
            \noalign{\smallskip}
            \hline
            \noalign{\smallskip}
SMC X-1 & 1.81(-0.71,+0.82) & 1.41(-0.41,+0.43) & 2.57   & 1.42   & 54  & 43 
& C \\
SMC X-2    & 10(-8.5)       & 2.93(-1.4,+7)     & 2.67   & 1.38   & 10  & 11 
& B1 \\
RX~J0032.9-7348 & 10(-9)    & 2.5(-1,+3)        & 0.25   & 0.13   & 20  & 19 
& F2 \\
RX~J0049.1-7250 & 1.1       & 13                & 2.5    & 1.6    & 1.5 & 5  
& B1 \\
                & 1.0       & 21                & 0.32   & 0.2    & 5.0 & 7  
& B2 \\
RX~J0051.8-7231 & 10(-8.5)  & 1.17(-0.4,+1.6)   & 0.509  & 0.350  & 45  & 49 
& E \\
RX~J0052.1-7319 & 5.4       & 3.34(-2,+17)      & 0.162  & 0.126  & 10  & 11 
& B2 \\
RX~J0101.0-7206 & 2.23(-1.5) & 1.63(-1.2,+3.4)  & 0.184  & 0.0804 & 17  & 18 
& A1 \\
            \noalign{\smallskip}
            \hline
            \noalign{\smallskip}
            \multispan{8}
 SMC~opacities \\
            \multispan{8} 
(galactic~contribution~of~0.3~$\rm \times$~$\rm 10^{21}~cm^{-2}$~subtracted) \\
            \noalign{\smallskip}
            \hline
            \noalign{\smallskip}
        & $\rm kT_{th}$ & $\rm N_H$                      & $\rm L_{x}$                & $\rm L_{abs}$     &        &           \\
 Source & [keV]     & [$\rm 10^{21} cm^{-2}$]        & [$\rm 10^{37} erg/s$]  & [$\rm 10^{37} erg/s$] & $\rm\chi_2$ & DOF & Field \\
            \noalign{\smallskip}
            \hline
            \noalign{\smallskip}
SMC X-1    & 1.94(-0.67,+2.71) & 2.38(-0.9,+0.98) & 1.77  & 1.34  & 49   & 43
& C \\
SMC X-2    & 10(-9)            & 5.70(-1.7,+12)   & 2.29  & 1.28  & 13   & 11
& B1 \\
RX~J0032.9-7348 & 10(-9)       & 5.3(-4,+8)       & 0.26  & 0.14  & 19   & 19
& F2 \\
RX~J0049.1-7250 & 8.0 & 33 & $\rm 8\times 10^3$ & $\rm 6\times 10^3$ & 1.5 &5
& B1 \\
          & 24                 & 41               & 1.3   & 1.0   & 5.1  & 5
& B2 \\
RX~J0051.8-7231 & 10(-8)       & 1.87(-1.1,+2.6)  & 0.492 & 0.349 & 54   & 49
& E \\
RX~J0052.1-7319 & 17           & 7.05             & 0.137 & 0.109 & 9.7  & 11
& B2 \\
RX J0101.0-7206 & 3.46(-3.46,+6.5)  & 2.46(-2)    & 0.152 & 0.0824 & 17  & 18
& A1 \\
            \noalign{\smallskip}
           \hline
           \end{tabular}
           \end{flushleft}
   \end{table*}

\begin{table}
  \caption[]{Hydrogen column densities for the spectrally soft and hard 
             candidate X-ray binaries as deduced from a spectral fit 
             assuming blackbody flux distributions for the supersoft sources 
             and thermal bremsstrahlung distributions for the hard sources. 
             The values for the hard sources have been taken from Table~6 and 
             for the supersoft sources from Kahabka et al. (1994). For 
             comparison hydrogen column densities, deduced from 21-cm radio 
             data (HI columns) are given. The latter values have been deduced 
             from a HI map reproduced from Luks (1994) and smoothed with a 
             Gaussian $\sigma$ of $\rm 8\amin$.}
  \begin{flushleft}
  \begin{tabular}{rccc}
  \hline
  \hline
  \noalign{\smallskip}
   Cat. & Source name     & \multicolumn{2}{c} {Hydrogen column}       \\
   num. &                 & \multicolumn{2}{c} {[$10^{21}\ cm^{-2}$]}  \\
        &                 & spectral fit         & 21-cm (HI)          \\
  \noalign{\smallskip}
  \hline
  \noalign{\smallskip}
   \multicolumn{4}{c} {hard X-ray sources}         \\
  \noalign{\smallskip}
  \hline
  \noalign{\smallskip}
   242 & SMC X-1         & 2.38(-0.9,+0.98)     & 2.89   \\
   100 & SMC X-2         & 5.70(-1.7,+12)       & 2.54   \\
     3 & RX J0032.9-7348 & 5.3(-4,+8)           & 0.3    \\
    69 & RX J0049.1-7250 & 33                   & 4.73   \\
    83 & RX J0051.8-7231 & 1.87(-1.1,+2.6)      & 2.79   \\
    84 & RX J0052.1-7319 & 7.1                  & 5.14   \\
   158 & RX J0101.0-7206 & 2.46(-2)             & 2.98   \\
  \noalign{\smallskip}
  \hline
  \noalign{\smallskip}
   \multicolumn{4}{c} {supersoft~sources}    \\
  \noalign{\smallskip}
  \hline
  \noalign{\smallskip}
    20 & 1E~0035.4-7230   & 0.5(-0.15,+0.3)      & 0.3    \\
   135 & 1E~0056.8-7146   & 0.77(-0.2,+0.8)      & 1.54   \\
    62 & RX~J0048.4-7332 & 5.8(-4.4,+1.1)       & 4.22   \\
   133 & RX~J0058.6-7146 & 0.28(-0.28,+6)       & 1.93   \\
  \noalign{\smallskip}
  \hline
  \end{tabular}
  \end{flushleft}
\end{table}

\subsection{All-Sky Survey Observations}

The field of the SMC (where X-ray binaries have been detected) has been
observed from about 21~October 1990 till 31~October 1990. Of the sample of 
spectrally hard X-ray binaries only SMC~X-1 has been detected during this 
observation and we briefly report about this observation. For the other 
hard X-ray binaries and candidate hard X-ray binaries only upper limit 
count rates can be given for this epoch. The observations of the supersoft 
X-ray binaries during the {\it RASS} have been reported in Kahabka et al. 
(1994).

\subsubsection{SMC X-1}

SMC~X-1 has been observed from 23~October 1990 (20:40~UT) till 28~October 
1990 (9:33~UT) and stayed in a low-state during the whole period with a mean
count rate of $\rm 0.595\pm0.074$ $\rm counts\ s^{-1}$ (Kahabka \& Pietsch 
1993). In Figure~8 (first panel) the light curve (with a binsize of 20~sec) 
is shown. It covers almost exactly a full binary orbital cycle of 3.89~days. 
The observation starts with the eclipse phase and enters then (during orbital 
phases 0.3-0.6) into a similar flaring phase as has been seen in the pointed 
observations (cf. Figure~8, second panel) but here with much lower statistical
significance. A thermal bremsstrahlung fit with two absorption components 
(a galactic contribution of $\rm 3\times10^{20}\ atoms\ cm^{-2}$ and a 
variable SMC intrinsic component with reduced metallicities, factor~7) applied
to the RASS spectrum of SMC~X-1 gives spectral parameters (kT$\rm \sim2$~keV, 
$\rm N_H$$\sim$$\rm 3\times10^{21}\ atoms\ cm^{-2}$) which are consistent 
(within the errors) with the parameters found for the pointed low-state 
spectrum.

\subsubsection{Upper Limit Count Rates for the not detected X-ray Binaries} 

The transients SMC~X-2 and SMC~X-3 have not been in outburst during the 
{\it RASS}. The observation time intervals and the upper limit count rates
of these two transients and of the other candidate X-ray binaries listed
in Table~3 are given in Table~5.

\begin{table}
  \caption[]{\ros all-sky survey $\rm 2\sigma$ upper limit count rates 
  (0.1-2.4~keV), observation intervals, and exposure times of the 9 candidate 
  X-ray binaries in the SMC (cf. Table~3) not detected in the all-sky survey
  observations.}
  \begin{flushleft}
  \begin{tabular}{lccc}
  \hline
  \hline
  \noalign{\smallskip}
Source name & $\rm 2\sigma$     & time                        & expos. \\
            & rate              & [UT 1990]                   & [sec]  \\
            & [$\rm 10^{-3}/s$] & start - end                 &        \\
  \hline
  \noalign{\smallskip}
SMC X-2         &  7.0          & 21 Oct 22:13 - 26 Oct 11:06 &  780 \\
SMC X-3         &  9.7          & 24 Oct 14:17 - 28 Oct 11:09 &  638 \\
RX J0032.9-7348 &  6.8          & 20 Oct 14:10 - 24 Oct 31:40 &  640 \\
RX J0049.1-7250 &  5.9          & 24 Oct 07:53 - 28 Oct 07:56 &  722 \\
RX J0051.8-7231 & 11.1          & 22 Oct 19:02 - 26 Oct 14:19 &  667 \\
RX J0052.1-7319 &  8.4          & 24 Oct 07:53 - 27 Oct 12:44 &  783 \\
RX J0059.2-7138 & 14.1          & 27 Oct 06:19 - 31 Oct 04:48 &  501 \\
RX J0101.0-7206 & 19.1          & 26 Oct 06:18 - 30 Oct 07:59 &  530 \\
  \noalign{\smallskip}
  \hline
  \end{tabular}
  \end{flushleft}
\end{table}

\subsection{Pointed observations}

We discuss in this section the new sample of SMC X-ray binaries, 
compiled after the extensive pointed observations with \ros . This 
should give a less biased view if compared to the limited sensitivity of 
the instrumentation before. The available \ros {\sl PSPC} pointed data 
towards the SMC, which have been analysed in this work allowed a by about 
a factor of $\sim$10 deeper search for candidate X-ray binaries than 
was possible with the {\it RASS} data. In Table~9 we give for those sources, 
which have been found during \ein and \ros observations the \ein and \ros 
name. We will discuss the individual SMC binary systems in detail. The 
complete list of known SMC X-ray binaries and candidate X-ray binaries 
is given in Table~10. 

\subsubsection{The high-mass binary system SMC X-1}

\begin{figure*}
  \centering{
  \vbox{\psfig{figure=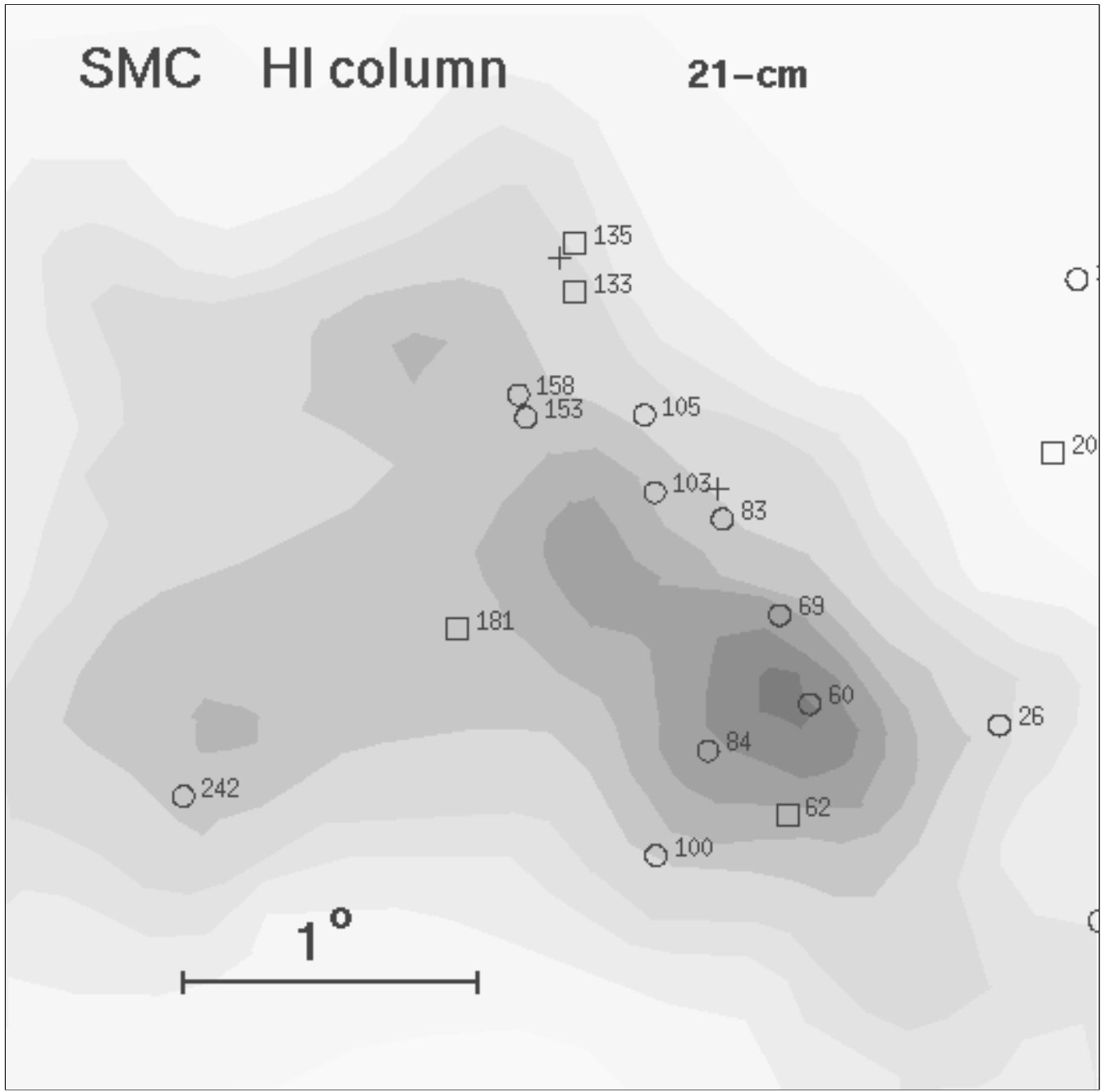,width=18.0cm,%
  bbllx=3.05cm,bblly=10.55cm,bburx=18.8cm,bbury=26.1cm,clip=}}\par
            }
  \caption{21-cm HI contour map (reproduced from Luks 1994) with the 
           positions of the candidate supersoft X-ray binaries (square) 
           and of the hard X-ray binaries (circle) drawn (cf. Figure~1).
           $\rm N_H$ contours are 70,60,50,40,30,20,15,10,5 $\rm \times10^{20}$
           hydrogen atoms $\rm cm^{-2}$. The darkest shading refers to the
           largest hydrogen column.} 
\end{figure*}

SMC~X-1 has been detected during a rocket flight (Price et al. 1971). The 
discovery of eclipses by the {\it UHURU} satellite with a period of 3.89~days 
(Schreier et al. 1972) established the binary nature of the source. The 
optical counterpart has been identified as a B0I supergiant (Sk~160) (Webster 
et al. 1972, Liller 1973). Optical photometry indicated the presence of an
accretion disk influencing the optical light curve (van Paradijs \& Zuiderwijk
1977). X-ray low- and high-intensity states have been observed (Schreier
et al. 1972, Tuohy \& Rapley 1975, Seward \& Mitchell 1981, Bonnet-Bidaud \& 
van der Klis 1981). Angelini et al. (1991) report about the discovery of an 
X-ray burst from SMC~X-1. This burst may be related to the type~II bursts 
observed from the Rapid Burster, which are interpreted as resulting from an 
instability in the accretion flow (Lewin, van Paradijs and Taam 1993). A 
$\sim$60~day quasi-periodicity in the X-ray flux has been suggested by Gruber
\& Rothschild (1984) from {\it HEAO-1 (A4)} data. Whitlock \& Lochner (1994) 
generated a 7~year light curve from {\it Vela~5} data which showed the source
only once in a prolonged high-state (of duration $\sim$100~days). Binary 
orbital parameters have been established by Primini, Rappaport \& Joss (1977). 
A decay in the binary orbit with $\rm \dot{P}_{orb}/P_{orb} = 
(-3.36\pm0.02)\times10^{-6}\ yr^{-1}$ was reported by Levine et al. (1993). 
They propose that this is caused by tidal interaction between the orbit and 
the rotation of the companion star (asynchronism) due to the evolutionary 
expansion of the companion star Sk~160 which is assumed to be in the Hydrogen 
burning phase.

During our \ros pointed observations covering one full binary orbit of 
SMC~X-1 the source was in a low intensity state. It was in the beginning 
observed during an extended and shallow eclipse (with residual flux at zero
orbital phase and a mean luminosity of $\rm \sim 8\times 10^{36} erg\ s^{-1}$).
It then passed into a flaring state with two additional bursts (separated by 
about 10~hours). The increase in intensity from the shallow eclipse to the 
peak of the first short duration flare is a factor of 10 and the peak 
luminosity in the flares is $\rm \sim 7\times 10^{37} erg\ s^{-1}$. This 
luminosity is deduced for a distance of 65~kpc. The true luminosity may be a 
factor of 2 lower, as from optical data a distance of $\sim$45~kpc is favored 
(Howarth 1982). This is consistent with SMC~X-1 being located in a spiral arm
extending from the bulge of the SMC towards our Galaxy (cf. Gardiner et al. 
1994). The spectrum of SMC~X-1 is soft with a thermal bremsstrahlung 
temperature close to 2~keV (cf. Figure~5). The X-ray eclipse of SMC X-1 has 
been found from COS-B data (2-12~keV) to have a duration of 0.65~days (cf. 
Bonnet-Bidaud \& an der Klis 1981). But in the \ros data we see an extended
and shallow eclipse with residual flux of $\rm \sim 0.1\ counts\ s^{-1}$ in 
the minimum. This may be explained by the fact, that the optical star appears
bigger in the soft X-ray band due to the stellar wind driven from its surface.

Bonnet-Bidaud \& van der Klis (1981) observed from COS-B data extending over 
an observation period of 38~days the transition from a low to a high intensity
state (a turn-on). They come to the conclusion, that the high-state luminosity
cannot be explained by wind accretion and propose Roche lobe overflow as 
origin. They propose quasi-periodic changes in the mass loss rate from Sk~160
e.g. due to g-modes (Papalaizou 1979) to be responsible for the switch over 
from low-intensity to high intensity states. The fact that SMC~X-1 was in a 
total of nine observations performed by {\it HEXE} (20-50~keV) only 4 times 
detectable - above a $\rm 5\sigma$ level - (Kunz et al. 1993), indicates 
that the low-states are not highly absorbed states of the X-ray source but 
more probably real changes in the accretion rate mediated by an accretion disk
to the NS. This is supported by the recently published 7~year light 
curve of SMC~X-1 generated from Vela~5B data (Whitlock \& Lochner 1994) in 
which the source only once has been found for about 100~days in a high 
intensity state.

SMC~X-1 has been observed by \ros (from 6-Oct-91 4:02~UT till 8-Oct-91 
3:05~UT) in a high-intensity state. This gives the unique opportunity to study
SMC~X-1 in two low-states separated by quite exactly 1~year and in one 
high-state which was observed shortly (only $\sim$7~days) before the second 
low-state. We therefore can study one transition in states (turn-off) in a 
similar way as has been possible for Bonnet-Bidaud \& van der Klis (1981) 
(with COS-B data) during a turn-on at higher energies (2-12~keV). We generated
the orbital light curve for the \ros high-state observation and show the 
result together with the (16-19)~October~91 low-state observation in Figure~8. 

\begin{figure}
  \centering{
  \vbox{\psfig{figure=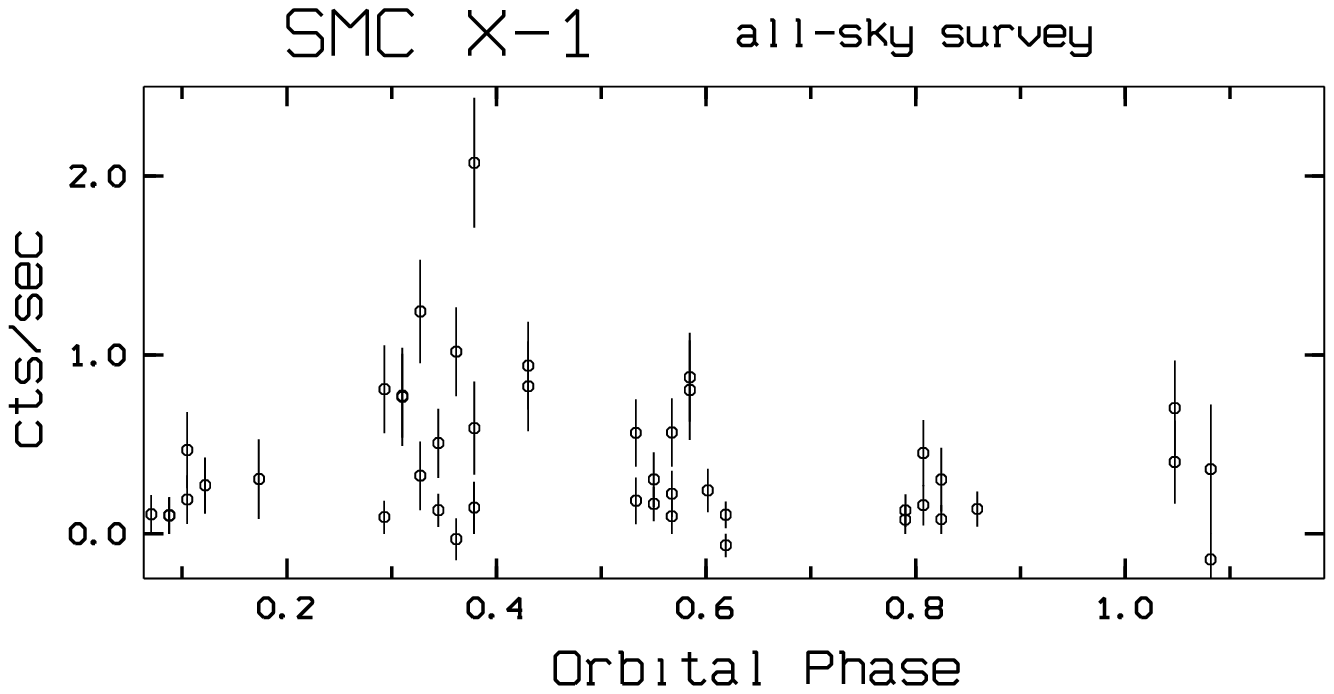,width=8.0cm,%
  bbllx=1.4cm,bblly=1.5cm,bburx=15.5cm,bbury=9.3cm,clip=}}\par
  \vbox{\psfig{figure=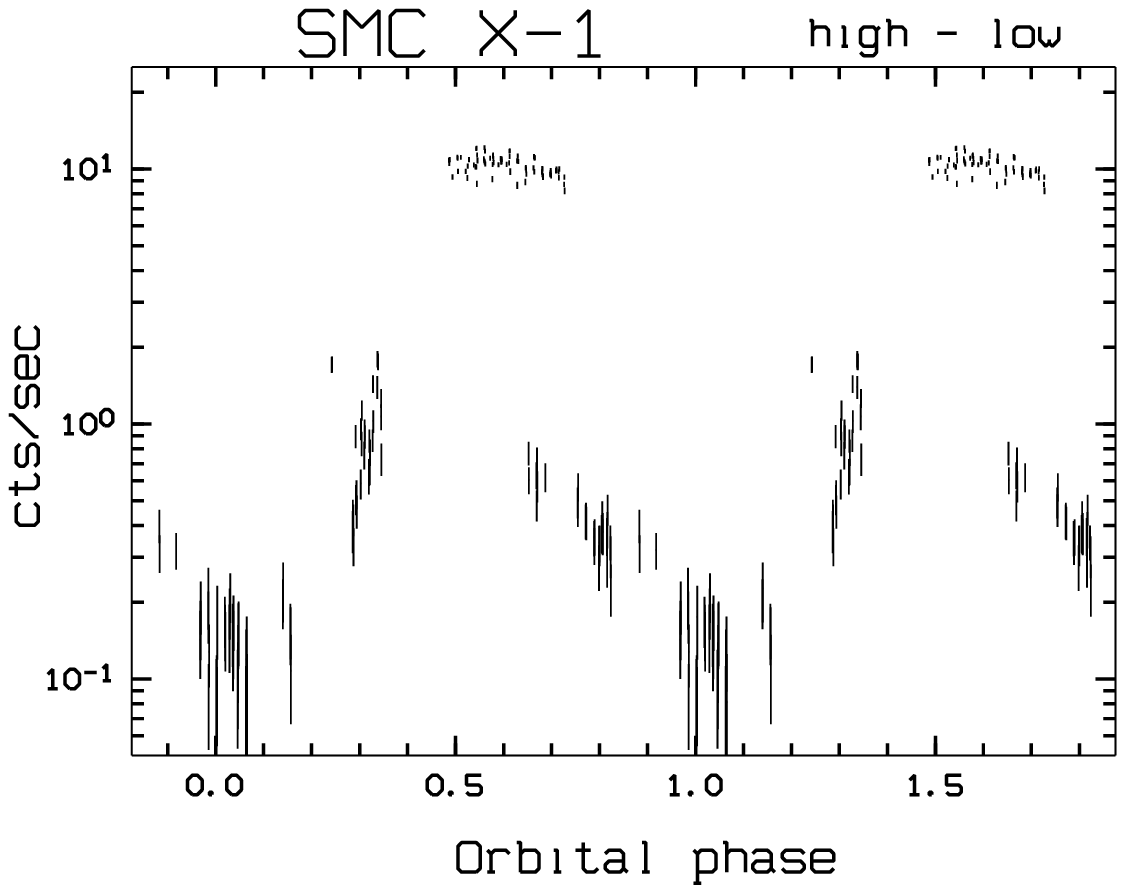,width=8.0cm,%
  bbllx=1.5cm,bblly=3.5cm,bburx=13.8cm,bbury=13.0cm,clip=}}\par
  \vbox{\psfig{figure=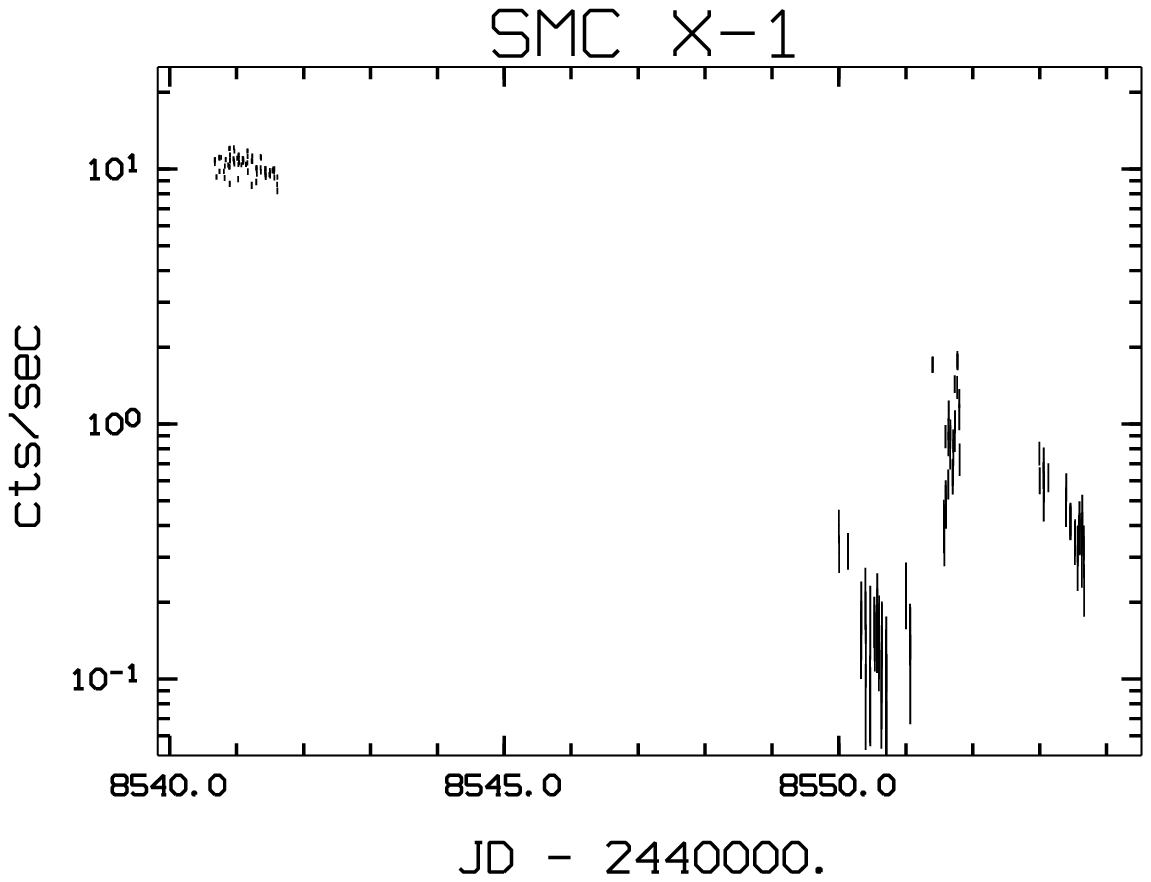,width=8.0cm,%
  bbllx=1.2cm,bblly=3.5cm,bburx=13.5cm,bbury=13.0cm,clip=}}\par
            }
  \caption{First panel: Binary orbit light curve (0.5-2.4 keV) of SMC X-1 as
           observed during the RASS. One data point is an integration of 
           20~sec. An orbital period of 3.89229188~days and an epoch for
           the mid eclipse center JD~2447740.85906 has been chosen (cf. 
           Levine et al. 1993).
           Second panel:  Binary orbit light curve (0.5-2.4~keV) of SMC~X-1 
           during a high and a low-intensity state. The high-state light curve
           is poorly sampled. Two cycles of the binary orbit are repeated.
           Third panel: Same data plotted as a function of time (Julian Date
           JD is given). The high-state precedes the low-state by about 
           7~days.}
\end{figure}

There is a clear spectral change observed with the source being much harder 
during the high-state indicating that the direct accretion luminosity from 
the NS is observed. This is supported by the fact, that X-ray pulsations have 
been discovered. The results of a detailed analysis of this high-state 
observation is presented in Blondin \& Woo (1995) and Woo et al. (1995). As 
the high-state observation covers only the phase region $\rm \sim0.5-0.7$ of 
the binary orbit, only a quite incomplete comparison is possible. The 
intensity in the high-state is a factor of $\sim$6 higher than in the two 
low-state flares.

\subsubsection{The Be type transient SMC X-2}

SMC~X-2 (2S~0052-739) is a transient X-ray source discovered by {\it SAS-3}
1977~October 11~to~16 (Clark et al. 1978). The X-ray luminosity (2-11~keV)
was deduced as $\rm 8.4\times10^{37}\ D_{65}^2$ erg/s, where $\rm D_{65}$ 
is the distance in units of 65~kpc. In a follow-up observation 1~month later 
the source was no longer detected. The $\rm 3\sigma$ upper limit of the 
luminosity was $\rm 8.4\times10^{36}\ D_{65}^2$ erg/s (Clark et al. 1979). 
This shows, that SMC~X-2 varies at least by a factor of 10 on a time scale 
of 1~month. It has not been detected a second time before \ros . An optical 
counterpart (within the 20'' error circle) proposed by Sanduleak \& Phillips 
(1977), has been studied by Allen (1977), van Paradijs (1977), Crampton et 
al. (1978), Murdin et al. (1979). The optical counterpart was resolved into 
two components of spectral type $\rm O7\pm2(V=15.2)$ and $\rm B1\pm2e(V=16.0)$
by Murdin et al. (1979). The association of X-ray sources with Be stars 
(Maraschi et al. 1976) led Murdin et al. (1979) propose the weaker component 
to be the more likely optical counterpart of SMC~X-2. Tarenghi et al. (1981) 
obtained a far UV spectrum of SMC~X-2 with the International Ultraviolet 
Explorer ({\it IUE}). They come to the conclusion that the O7 star dominates 
the UV spectrum and not the B1e star. The question remains open which 
component is the optical counterpart. 

\ros captured SMC~X-2 during another X-ray outburst 14 years after the first 
observed outburst in (1977). The observation lasted only for 1350~seconds. The 
source had a luminosity of $\rm 2.7\times 10^{37} erg\ s^{-1}$ (0.15-2.4 keV) 
taking into account a column density of $\rm 2.9\times 10^{21} atoms\ cm^{-2}$
and a thermal bremsstrahlung temperature of 10~keV from the spectral fit. The
value for the absorption has been deduced by assuming a galactic contribution 
of $\rm 3\times 10^{20}\ cm^{-2}$ and a SMC intrinsic contribution of 
$\rm 2.1\times 10^{21}\ cm^{-2}$ with metals reduced by a factor of 
$\rm \sim 7$. The luminosity observed with \ros (0.15-2.4 keV) is a factor 
of 3 
smaller than the luminosity observed by {\it SAS~3} (2-11 keV). The difference
in energy band cannot account for this difference as \ros may have seen the 
higher luminosity part of the spectrum. Since neither the \ros nor the 
{\it SAS~3} observation covered the rise and decay of the outburst it is not 
clear, if the luminosity difference reflects a difference in the peak 
luminosities of the two observed outbursts. 

\subsubsection{The binary candidate RX~J0051.8-7231}

The earlier \ein error box of RX~J0051.8-7231 of $\rm 40\asec$ radius (source
\#27 in the {\it Einstein} catalog of Wang \& Wu 1992) coincides with the 
bright SMC B1 star AV~111 (Bruhweiler 1987, Wang \& Wu 1992). But within the 
smaller error circle deduced from the \ros observations with a radius of 
$\rm 11\asec$ a blue emission line star has been found, which is a likely 
optical candidate (Pakull, private communication). The source flux increased 
by a factor $\rm >10$ over a period of $\sim$1~year and by a factor of $\sim$5
in 5~days with a significant variation in the hardness ratio. The moderate 
X-ray luminosity ($\rm 10^{35}-10^{38}\ erg\ s^{-1}$), the large flux 
variation, the relatively hard spectrum and the early-type star counterpart 
suggested a Be type origin of the X-ray source. Recently Israel et al. (1995)
reported about the discovery of 8.9~s pulsations in \ros data of this source
(in pointing E, cf. Table~1). This finding makes the source a firm X-ray 
binary in the SMC. An orbital period has not yet been determined. 

From \ros observations we confirm RX~J0051.8-7231 (WW~27) to be variable by a 
factor $>$10. During a pointing covering 3~days the source showed a nearly 
linear increase in luminosity ranging from $\rm \sim 2\times 10^{36} erg\ 
s^{-1}$ to $\rm \sim 6\times 10^{36} erg\ s^{-1}$ deduced from a spectral fit 
with a SMC intrinsic absorption of $\rm 2\times 10^{21} atoms\ cm^{-2}$ (cf. 
Figure~3). In two other pointings the luminosity was $\rm \le 1.1\times 
10^{35} erg\ s^{-1}$ (cf. Table~3 and Table~5). This behavior is 
characteristic for HMXBs with luminosities below $\rm 10^{37} erg\ s^{-1}$ 
showing erratic flaring activity (cf. White 1989).

\subsubsection{The binary candidate RX~J0052.1-7319}

RX~J0052.1-7319 has already been discovered in \ein observations (source \#29 
in the catalog of Wang \& Wu (1992) and \#8 in the catalog of Inoue, Koyama \&
Tanaka (1983)). It correlates with the nebular complex DEM~70
(Davis, Elliot \& Meaburn 1976). The source has not been considered as an 
X-ray binary candidate from \ein observations. \ros detects this source to be 
variable by a factor of $\sim$4 in an observation (B2) covering 5 days (cf. 
Figure~4). The X-ray luminosity varies from $\rm 6\times 10^{35} erg\ s^{-1}$ 
to $\rm 2.6\times 10^{36} erg\ s^{-1}$ deduced from a spectral fit with a SMC 
intrinsic absorption of $\rm 7\times 10^{21} cm^{-2}$. An optical counterpart
has not been found for this source. Kahabka (1995a) gives a correlation with 
a B star for this source. But this is in error.

\begin{table}
  \caption[]{\ein and \ros names of those X-ray binary sources, which have
             been found during \ein and \ros observations. The \ein name 
             is adopted or deduced (from the catalogued coordinates) from
             the information given in Wang \& Wu (1992). For the two supersoft
             sources 1E~0035.4-7230 and 1E~0056.8-7146 we did not introduce a 
             new \ros name, as these sources have been optically identified 
             before the \ros observations.}
  \begin{flushleft}
  \begin{tabular}{cc}
  \hline
  \hline
  \noalign{\smallskip}
    \ein name      &  \ros name   \\
  \noalign{\smallskip}
  \hline
  \noalign{\smallskip}
    1E~0035.4-7230  & --              \\
    1E~0050.1-7247  & RX~J0051.8-7231 \\
    1E~0050.3-7335  & RX~J0052.1-7319 \\
    1E~0056.8-7146  & --              \\
  \noalign{\smallskip}
  \hline
  \end{tabular}
  \end{flushleft}
\end{table}

\subsubsection{Discovery of a new transient RX~J0101.0-7206}

RX~J0101.0-7206 has been discovered by \ros in pointing A1 at a luminosity of  
$\rm 1.3\times 10^{36} erg\ s^{-1}$. About half a year later the X-ray source 
has disappeared with a $\rm 2\sigma$ upper limit X-ray luminosity (assuming 
that the spectral parameters did not change) of $\rm 4.6\times 10^{34}\ erg\ 
s^{-1}$ (cf. Table~3 and Table~5). The source has not been detected in \ein 
observations. This suggests a transient nature. The source most probably is 
associated with a 15-16`th magnitude Be star (Pakull, private communication). 
RX~J0101.0-7206 lies at the north-eastern boundary of the star formation 
region N66. A few additional but persistent X-ray sources have been detected 
by \ein and confirmed with \ros clustering around N66 (Kahabka et al. 1996). 
They have been interpreted as SNRs (Ye et al. 1991).

\subsubsection{RX~J0049.1-7250, an AGN or X-ray binary}

RX~J0049.1-7250 has been discovered with \ros in pointing B1. It is extremely 
absorbed with $\sim$1-3$\rm \times 10^{22} atoms\ cm^{-2}$ (cf. Table~6) and 
most probably lies at the far side of the SMC seen through the bulge (with a 
total HI column of $\rm \sim5.5\times10^{21}\ cm^{-2}$, cf. Figure~7). The 
best-fit thermal bremsstrahlung temperature of 1~keV gives a luminosity of 
$\rm 2.5\times 10^{37} erg\ s^{-1}$ for using galactic absorbing opacities. 
A luminosity above $\rm 10^{38} erg\ s^{-1}$ is deduced for SMC opacities. 
But the uncertainty in the column density determination is probably too high 
to assume that the source was in a super-Eddington bright state. Half a year 
later the luminosity has decreased by a factor of 10 or more with the 
absorption not markedly changed. This confirms the source to be behind the 
SMC probably seen during and after a flare. We cannot at the moment exclude 
that RX~J0049.1-7250 is a time variable AGN shining through the SMC. If the 
source is a SMC X-ray binary then it could be located in the far-off spiral 
arm (wing) of the SMC (see the simulation of the Magellanic system by Gardiner
et al. 1994). In the bright state it would show a similar X-ray luminosity 
as SMC~X-1, which is supposed to lie in the spiral arm pointing towards our 
Galaxy.

\subsubsection{RX~J0032.9-7348}

This source lies at the south eastern boundary of the body of the SMC. It
is only contained in pointings F1 and F2. The source is variable by a factor
of $\sim$6 over a time of $\sim$0.4~years. In the X-ray bright state (pointing
F2) a spectral fit could be applied. An absorbing hydrogen column of 
$\rm 2.5\times10^{21}\ atoms\ cm^{-2}$ and $\rm 5.3\times10^{20}\ atom\ 
cm^{-2}$ is obtained in a thermal bremsstrahlung and in a blackbody 
description respectively. A temperature of $\rm >1~keV$ and 0.6~keV is 
obtained. The unabsorbed bolometric blackbody luminosity would be 
$\rm 2.5\times10^{36}\ erg\ s^{-1}$ and the thermal bremsstrahlung luminosity 
(0.1-2.4~keV) $\rm 2.6\times10^{36}\ erg\ s^{-1}$ for a distance of 65~kpc 
(SMC membership). Assuming that the source spectrum did not change from 
pointing F1 to F2, an (unabsorbed) luminosity of $\rm 3\times 10^{35}\ erg\ 
s^{-1}$ is deduced for the X-ray faint state. A 15.3~mag blue object, 
GSC~0914101338 is $\rm 22\asec$ from the X-ray position and may be a good 
candidate (Pakull, privte communication). The star HV~1328 has been found 
in the {\sl Strasbourg catalog} at a distance of $\rm 67\asec$ from the 
source. This is about the positional uncertainty at the large off-axis 
angle ($\rm 43\amin$) where the source has been found. As the X-ray spectrum 
cannot be explained by coronal emission and as the duration of the X-ray 
bright state is $\rm \sim$5~days (cf. Figure~4) a flare star nature may be 
excluded. These facts are in favor for a SMC membership of the source. The 
hard spectral characteristic is in favor for a HMXB nature of the source.

\subsubsection{SMC X-3}

SMC~X-3 (2S~0050-727) has been discovered with {\it SAS~3} at a luminosity 
(2-11~keV) of $\rm 5.9\times10^{37}\ D_{65}^2\ erg/s$ (Clark et al. 1978).
It is a transient source probably of the Be type. In a follow-up 
observation 1~month later the source was no longer detected (the $\rm 3\sigma$ 
upper limit for the luminosity is $\rm 8.3\times10^{36}\ D_{65}^2\ erg/s$). 
The source varies at least by a factor of 7 within 1~month (Clark et al. 
1979). No second detection of the source has been reported up to now.

SMC~X-3 has been in the field of view of the pointings A1, A2 and E (cf. 
Table~5). It has not been detected in these observations. The upper limit to 
the count rates (cf. Table~5) indicates to the very low luminosity (below 
$\rm \sim 3.5-10\times 10^{34}\ erg\ s^{-1}$, assuming a SMC~X-2 like 
spectrum) during these periods. This is in agreement with the Be type 
transient identification of SMC~X-3. We also refer to the paper of Pietsch et 
al. (1986), where an upper limit luminosity of $\rm 2\times10^{33}\ erg\ 
s^{-1}$ (2-6~keV) has been deduced for the Be type transient GX~304-1 from 
{\it EXOSAT} observations around the time of an expected periodic outburst. 
It has been argued, that the shell of the Be system has diminished in this 
system.

\subsubsection{The 2.8~s transient RX~J0059.2-7138}

This transient has been discovered by Hughes (1994) as a serendipitous source 
in a pointed observation on 12~May~1993 towards the bright supernova remnant 
SNR~0102-72.2 in the SMC. The transient was very bright ($\rm 7.8~cts\ counts\ 
s^{-1}$) and showed a pulsed signal of 2.76~s. The source is probably 
correlated with a blue $\rm B_j\sim 14.1\ mag$ HST guide star. As the source 
has not been seen in previous observations of the SMC (neither \ein nor \ros) 
Hughes concluded the source is probably a Be type transient. We add in our 
analysis further pointed observations towards the direction of this pulsar, 
however did not detect the source in an observation 44~days before the
reported outburst nor 142~days afterwards. This limits the maximum outburst 
duration to less than 6~months. Upper limits for the source are more than a 
factor of 1000 below the outburst luminosity (cf. Table~5).

\subsubsection{H~0107-750}

H~0107-750 (1H~0103-762) has been seen with the {\it HEAO A1} (Wood et al. 
1984) and {\it HEAO A3} (Tuohy et al. 1988) experiment. Whitlock \& 
Lochner (1994) found in {\it Vela~5B} data mapping the region of the source 
for 7~years three outbursts with luminosities of $\rm \sim4\times10^{38}$, 
$\rm \sim4\times10^{38}$ and $\rm \sim6\times10^{38}\ erg\ s^{-1}$ in the 
3-12~keV energy band. The outburst occurred in 1969~July, 1969~October and 
1970~February. The outburst lasted for $\sim$35 days.

The position of H~0107-750 is not covered by any of our pointed \ros
observations. The source has not been detected during the RASS (cf. Kahabka
\& Pietsch 1993).

\section{Discussion}

\subsection{Transient and persistent sources}

It is of interest to compare the number of persistent with the number of 
transient X-ray binary candidates. A source is considered to be a transient, 
if it is in the field of view of at least two observations and below the 
detection limit of at least one observation. From Table~5 it becomes clear 
that four of the spectrally hard sources, e.g. SMC~X-2, RX~J0101.0-7206, 
SMC~X-3, RX~J0059.2-7138 fall into this category. They have not been detected 
in least one pointing with a $\rm 2\sigma$ upper limit count rate in the 
0.5-2.4~keV band of $\rm <7.\times10^{-4}\ counts\ s^{-1}$. This corresponds 
to a luminosity below $\rm \sim2.\times10^{34}\ erg\ s^{-1}$. They all are (or
may be) Be type transients. Two sources, RX~J0051.8-7231 and RX~J0052.1-7319, 
have not been detected at least once. But both sources are known from \ein 
observations and we classify them as persistent and variable. RX~J0049.1-7250 
and RX~J0032.9-7348 are also considered as persistent and highly variable. 
This gives, including SMC~X-1, 5 persistent (and variable) and 4 transient 
hard X-ray binary candidates. 

The supersoft X-ray binary sample (with a much smaller number of objects) 
gives two persistent sources (1E~0035.4-7230 and 1E~0056.8-7146) and two 
transient sources (RX~J0048.4-7332 and RX~J0058.6-7146). In the supersoft
sources, which are in a steady-state nuclear burning condition, variability 
in the X-ray flux (by a factor in excess of $\approxlt$3) is not predicted 
to occur (cf. Fujimoto 1982). The recurrently burning systems (cf. Fujimoto 
1982, Kahabka 1995b) show variability. Variability in the X-ray flux related 
to the binary orbit is also observed in a few systems (like 1E~0035.4-7230, 
cf. Kahabka 1996). 

In our Galaxy the existence of $\sim$40 HMXB systems (with evolved companions)
and of $\rm (2-6)\times10^3$ Be type X-ray binaries has been estimated by van 
Paradijs \& McClintock (1995). This clearly shows that the Be type transients
outnumber the persistent and variable sources by a substantial number and 
ever more transients are expected to be discovered in future observations 
(there may be presently $\sim$300 Be type transients in the SMC, cf. Kahabka 
1995a). 

A mechanism for transient behavior different from that of the Be star 
phenomenon has been proposed by Stella, White \& Rosner (1986), centrifugal 
inhibition of accretion (see also Waters and van Kerkwijk 1989). Stella et al.
(1989) have shown, that a fair fraction of the NSs in transient X-ray binaries 
containing OB supergiant secondaries rotate close to the critical equilibrium 
rate at which the corotation radius equals the magnetospheric radius. A 
moderate increase in the mass accretion rate can turn these sources (during 
outburst) into very luminous transients. Otherwise these sources may be 
dormant. For a magnetic dipole moment of $\rm \sim10^{30}\ Gauss\ cm^{-3}$ and
(maximum) X-ray luminosities of $\rm \sim10^{36}-10^{38}\ erg\ s^{-1}$ NS 
rotation periods of $\sim$0.7-50~s are predicted if the system is close to 
centrifugal inhibition. For typical wind parameters of supergiant and Be X-ray
binaries orbital periods of $\sim$1-40~days are expected. SMX~X-1 falls well 
into this regime. Probably part of the X-ray variability we observe in the 
persistent sources may be related to this phenomenon.  
 
\subsection{The luminosity distribution}

For the seven hard X-ray (binary) sources in the SMC a number-luminosity 
diagram has been generated. The values for the luminosity (0.15 - 2.4 keV) 
were taken from Table~6. SMC~X-1 was observed in a low luminosity state 
(cf. Marshall et al. 1983), RX~J0059.2-7138 has not been included in the 
diagram as the high (bolometric) luminosity reported for the soft component 
is based on a blackbody fit and may be doubted or is at least highly 
uncertain. SMC~X-3 has not been seen with \ros and has not been included 
either. The transient X-ray sources have luminosities (0.15-2.4~keV) ranging 
from $\rm 5\times 10^{35} erg\ s^{-1}$ to $\rm \sim 10^{38} erg\ s^{-1}$ (cf. 
Figure~9). These luminosities are considerably lower than the super-Eddington 
luminosities inferred for SMC~X-1, SMC~X-2 and SMC~X-3 (Clark et al. 1978). 
But one has to note, that with \ros SMC~X-1 has been found during a high-state
in a super-Eddington bright state (Woo et al. 1995) and for the new discovered
SMC transient RX~J0059.2-7238 a supersoft component with bolometric 
super-Eddington luminosities has been measured (Hughes 1994). This is in favor
for the luminosity distribution of the SMC HMXBs extending to large values,
probably to larger values as is found in the galactic supergiant and Be type 
X-ray binaries (cf. van Paradijs \& McClintock 1995). In the 1970's, when the 
very luminous MC X-ray sources were discovered, the reduced metallicities of 
the Clouds have been considered as a possible link to a luminosity increase. 
X-ray heating (of a not fully ionized gas) depends due to photoabsorption 
strongly on the atomic number Z (cross-section for photoabsorption 
$\rm \sigma_{ph}\propto Z^4$). A lower Z gas is less affected by heating and 
higher accretion luminosities can be reached in agreement with the observation
of higher luminosity sources in the MCs (cf. discussion in van Paradijs \& 
McClintock 1995). Obviously we now observe with \ros in addition to these 
super-Eddington bright sources (due to the increased sensitivity) the 
lower-luminosity distribution. One can now ask the question whether the whole
luminosity distribution of the MC HMXBs is just shifted to larger luminosities
or extends over a broader range. This question cannot be answered yet. But the
observations indicate, that the outburst luminosities of the (identified and 
classified) MC Be-type X-ray binaries are well above $\rm 10^{35}\ erg\ 
s^{-1}$, the lower limit found for the galactic pendants (Figure~2 of van 
Paradijs \& McClintock 1995).       

\begin{figure}
  \centering{
  \vbox{\psfig{figure=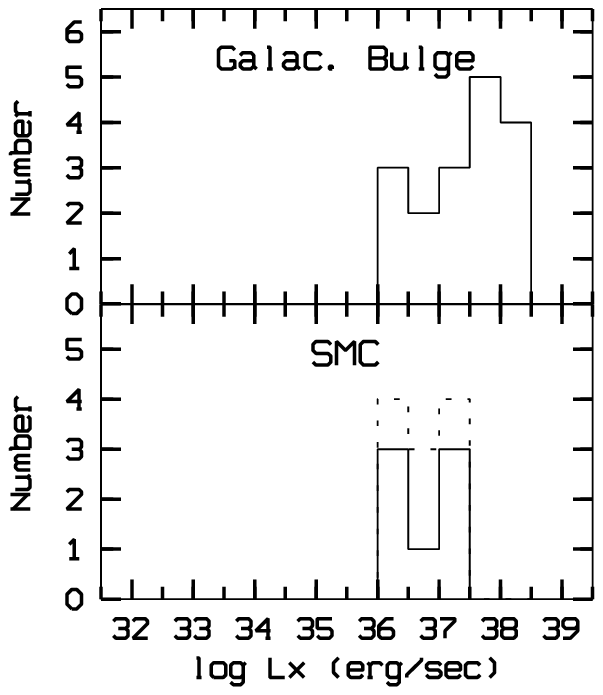,width=8.5cm,%
        bbllx=2.5cm,bblly=4.5cm,bburx=9.5cm,bbury=9.0cm,clip=}}\par
            }
  \caption[]{Number-luminosity diagram of the SMC X-ray binaries with 
             luminosities in the energy range 0.15-2.4~keV. The solid line 
             gives the distribution of the hard X-ray binaries and the 
             dotted line the distribution with the supersoft sources added.}
\end{figure}

\subsection{Extrapolation from the SMC to the Galaxy}

With the SMC we have due to its closeness to the Galaxy the unique 
opportunity to study stellar populations outside of our Galaxy. With the 
deep \ros pointings we covered a large fraction of this galaxy rather
homogeneously. Using a point source detection procedure and applying 
selection criteria for a candidate sample of X-ray binaries we were able 
to set up a rather firm sample with an intrinsic source luminosity above 
$\rm 3\times10^{35}\ erg\ s^{-1}$. This gives in total 10 persistent and
transient hard X-ray binaries and 4 supersoft binaries. Scaling with the 
mass ratio between the SMC and the Galaxy of $\rm {1\over 60}$ (McGee \&
Hindman 1971) $\sim$400 hard X-ray binaries and $\sim$250 supersoft X-ray
binaries would be expected to be seen in the Galaxy in case the star 
formation rate in the SMC and in our own Galaxy were similar (cf. for a 
discussion Clark et al. 1978, see also introduction). As the metallicity 
of the SMC is strongly reduced one can assume that there are not many 
hidden sources and this extrapolation is close to the total number of 
active hard and supersoft sources. The number of 250 SSS would be reasonably 
close to the number of $\sim$1000 expected galactic supersoft sources 
(Rappaport et al. 1994). van den Heuvel et al. (1992) estimated the number 
of HMXBs in the Galaxy containing a NS to $\rm \sim 10^4$. Obviously only a 
small fraction of them is active with luminosities above a few $\rm 10^{35}\ 
erg\ s^{-1}$. The total number of known (steady and transient) X-ray 
sources in the Galaxy is $\approxlt100$ (van Paradijs 1995). The latter 
discrepancy could be explained by a better sampling of the SMC compared 
to the Galaxy.

\subsection{The nature of the rejected X-ray binary candidates}

Certainly our selection criteria for X-ray binaries are not giving a firm 
sample of candidates. Nor can we be sure, that any of the rejected candidates
(cf. Table~3) will not turn out to be an X-ray binary in the SMC. This
decision may only be secure, after an optical identification of the objects 
in debate has been achieved. The X-ray nature of the sources not discussed in 
this paper will be the subject of a forthcoming paper. A first spectral and 
timing analysis of the rejected candidates listed in Table~2 with off-axis 
angles $\rm <30\amin$ has been performed. Only the source RX~J0100.7-7206 has 
the spectral appearance of a background AGN, e.g. a large SMC intrinsic 
hydrogen column (assuming metallicities reduced by a factor of $\sim$7) of 
$\rm 3.1\times10^{21}\ cm^{-2}$ and a powerlaw photon index of -2.3 (cf. 
Figure~10). The value of the hydrogen column is consistent with the total SMC 
HI column in this direction deduced from 21-cm data (Luks 1994, cf. 
Figure~7). A flux (in the 0.1-2.4~keV band) of $\rm 3.12\times10^{-13}\ erg\ 
cm^{-2}\ s^{-1}$ is deduced. It may be interesting to note that Ye \& Turtle 
(1993) found in their 843~MHz radio survey of the SMC no significant 
difference in the number of background sources (e.g. AGNs) compared to regions
distant from the Clouds and at high Galactic latitudes.

For the other two sources RX~J0054.9-7226 and RX~J0055.4-7210 low SMC 
intrinsic hydrogen column densities of $\rm 2.7\times10^{20}$ and 
$\rm 1.3\times10^{21}\ cm^{-2}$ have been found and power law indices of
-0.2 and -2.1. Taking into account the uncertainties in the deduced parameters
then RX~J0055.4-7210 may still be compatible with an AGN. The low absorbing 
hydrogen column found for RX~J0054.9-7226 would indicate either a galactic 
foreground or a SMC object at the near side of the SMC.

\begin{figure}
  \centering{
  \vbox{\psfig{figure=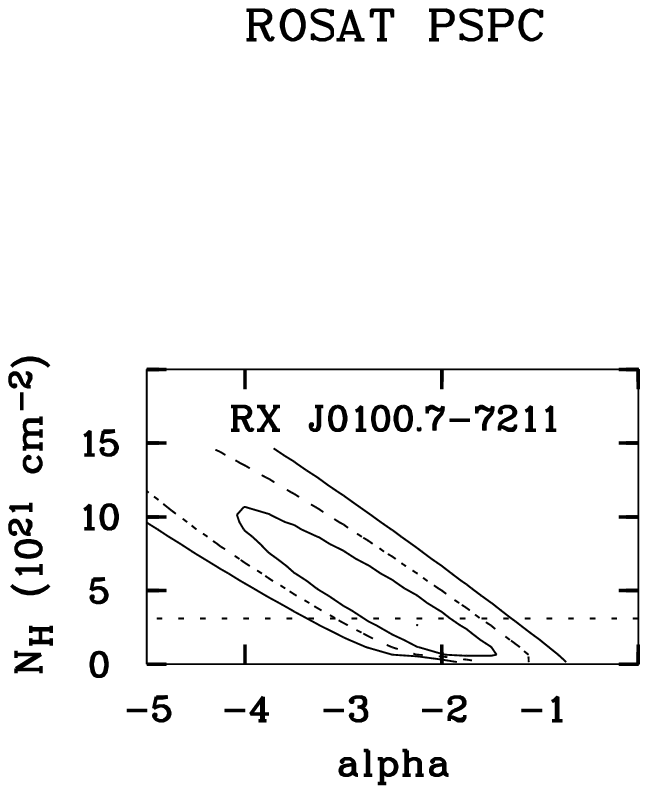,width=9.0cm,%
  bbllx=9.4cm,bblly=13.6cm,bburx=16.5cm,bbury=18.5cm,clip=}}\par
            }
  \caption{68, 95, and 99\% confidence parameter plane for the
  powerlaw photon index versus the hydrogen absorbing column density 
  within the SMC (assuming metal abundances reduced by a factor of 
  $\sim$7 compared to galactic abundances) for the possible background
  AGN RX~J0100.7-7211, shining through the SMC bulge with a total
  SMC hydrogen column of $\rm 3.1\times10^{21} cm^{-2}$ (dashed line, cf.
  Luks 1994 and Figure~7).}
\end{figure}

   \begin{table*}[htbp]
      \caption[]{Spectrally hard and supersoft X-ray binaries (and candidate
                 X-ray binaries) in the SMC. The following information is 
                 given: The name of the source, the \ros {\sl PSPC} count 
                 rate, the temperature of the soft blackbody component
                 (given also for the hard binaries in case an additional soft 
                 component has been detected or has been known before), the 
                 temperature of the hard thermal bremsstrahlung component,
                 the optical counterpart, the visual optical magnitude,
                 the orbital period, the spectral type of the optical 
                 counterpart, remarks and references. 
                }
            \begin{center}
            \begin{tabular}{clrcccccccc}
            \hline
            \hline
            \noalign{\smallskip}
Name        & PSPC  & \multicolumn{2}{c}{kT}  & variab.& opt. ID & $\rm m_v$ & type & orb.Per. & remarks & Ref.$\rm ^{a}$ \\
            & [c/s] & soft & hard             &        &         &       &    
 & [d]      &         &      \\
            &       & [ev] & [keV]            &        &         &       &      &          &         &           \\  
            \noalign{\smallskip}
            \hline
            \noalign{\smallskip}
\multicolumn{11}{c}{Hard~X-ray~Binaries~(\&~candidates)} \\
            \hline
            \noalign{\smallskip}
SMC X-1        & 0.37  & -   & $\rm 1.8^{-0.7}_{+2.7}$ & flares   & Sk 160 
& 13.2    & B0I & 3.89   & low-state & 1-20 \\ 
               &       & 160 &                         & persist. &        
&       &     &          & high-state &   \\
SMC X-2        & 0.39  & -   & $\rm 10^{-9}_{+0}$      & trans.   & +      
& 15.2/ & $\rm O7\pm2/$  & -  & rediscovered    & 21-30 \\
               &       &     &                         &          &         
& 16.0  & $\rm B1\pm2e$  &    &                 &       \\    
SMC X-3        & -     &     &                         &          &        
&       & $\rm O9 (III-V)e$ &        &                     & 22-23 \\
RXJ0032.9-7348 & 0.12  & -   & $\rm 10^{-9}_{+0}$      & variab.  & ?      
& -    &        & -      & candidate    &    \\ 
RXJ0049.1-7250 & 0.048 & -   & 1-8                     & variab.  & -      
& -     &      & -      & candidate    & 21 \\
RXJ0051.8-7231 & 0.11  & -   & $\rm 10^{-8}_{+0}$      & variab.  & ?      
& -     &      & -      &        & 21,31-32,42 \\
RXJ0052.1-7319 & 0.021 & -   & 1-17                    & variab.  & -      
& -     & -    & -      & candidate    & 21 \\   
RXJ0059.2-7138 & 7.8   & 35  & ?                       & trans.   &        
& 14.1  & blue & -     &  candidate    & 41 \\
RXJ0101.0-7206 & 0.048 & -   & $\rm 3.5^{-3.5}_{+6.5}$ & trans.   & -      
& -        & -   & -      & candidate    & 21 \\
H0107-750      &       &     &                         & variab.  & -      
&     -  & Be  & -      & candidate & 40 \\       
           \noalign{\smallskip}
           \hline
           \noalign{\smallskip}
\multicolumn{11}{c}{Supersoft~X-ray~Binaries~(\&~candidates)} \\
           \noalign{\smallskip}
           \hline
           \noalign{\smallskip}
1E~0035.4-7230  & 0.38  & 41 & -                  & variab.& +       & 20.2  
&      & 0.17  &                 & 31,33-35 \\
1E~0056.8-7146  & 0.33  & 28 & -                  &           & N67  & 16.6  
& PN   & -     &                 & 33,37-39\\
RXJ0048.4-7332 & 0.19  & 20 & -                  &           & SMC3 & 15.5  
& symb.nova   & -     &                 & 33,36 \\
RXJ0058.6-7146 & 0.025 & 42 & -                  & outburst  & -    & -     &      & -     &                 & 33 \\
              \noalign{\smallskip}
           \hline
           \end{tabular}
           \end{center}
$\rm ^{a)}$ Ref.: \\
 (1) Angelini, Stella \& White, 1991;
 (2) Bonnet-Bidaud \& van der Klis, 1981;
 (3) Bonnet-Bidaud, et al., 1981;
 (4) Bunner \& Sanders, 1979;
 (5) Darbro, 1981;
 (6) Davison, 1977;
 (7) Hammerschlag-Hensberge, et al., 1984;
 (8) Henry \& Schreier, 1977;
 (9) Howarth, 1982;
(10) Hutchings \& Crampton, 1977;
(11) Kunz et al., 1993;
(12) Levine, et al., 1993;
(13) Marshall, White \& Becker, 1983;
(14) Osmer \& Hiltner, 1974;
(15) Primini, Rappaport \& Joss, 1977;
(16) Reynolds, et al., 1993;
(17) van Genderen, 1974;
(18) van Genderen \& van Groningen, 1981;
(19) van der Klis, et al., 1982;
(20) Wilson \& Wilson, 1976;
(21) Kahabka 1995;
(22) Clark et al. 1978;
(23) Clark et al. 1979;
(24) Sanduleak \& Philips 1977;
(25) Allen 1977;
(26) van Paradijs 1977;
(27) Crampton et al. 1978;
(28) Murdin et al. 1979;
(29) Maraschi et al. 1976;
(30) Tarenghi et al. 1981; 
(31) Wang \& Wu 1992;
(32) Bruhweiler 1987;
(33) Kahabka et al. 1994;
(34) Orio et al. 1994;
(35) Schmidtke et al. 1994;
(36) Vogel \& Morgan 1994;
(37) Wang 1991;
(38) Wang \& Wu 1992;
(39) Heise et al. 1994;
(40) Whitlock \& Lochner 1994;
(41) Hughes, 1994;
(42) Israel et al. 1995.
   \end{table*}
%

\section{Summary}

A systematic search for spectrally hard and soft X-ray binary systems in 
the SMC has been performed in the \ros {\sl PSPC} pointed data towards 
this galaxy. A selection has been applied to a list of X-ray sources to 
find candidate X-ray binaries. Detectable time variability 
and spectral properties determined from hardness ratio arguments were used 
as the criteria. The luminosity threshold was $\rm \sim3\times10^{35}\ erg\ 
s^{-1}$ mainly depending on the assumed absorbing column density towards the 
source. A catalog search for optical candidates coinciding with the error 
circle of the X-ray source has been performed and used as an identification 
criterion. From a list of 15 candidate hard X-ray binaries and 5 
candidate soft X-ray binaries 7 and 4 respectively were finally selected 
after taking into account other information like correlation with a SMC O or 
B star. A few selected sources can still be (time variable) background AGNs 
which have a chance superposition with the SMC. The supersoft sample has been 
discussed in Kahabka et al. (1994). The most prominent X-ray source is 
SMC~X-1. It was until recently the only SMC X-ray binary for which an orbital 
period has been determined. This situation changed, when a $\rm 4.^h1$ orbital 
period has been discovered in the optical data of the supersoft SMC source 
1E~0035.4-7230 (Schmidtke et al. 1994). SMC~X-1 is also one of the few (2) SMC 
NS X-ray binaries for which the rotation period of the NS has been determined 
(besides RX~J0059.2-7138 and RX~J0051.8-7231). SMC X-1 is known to show low 
and high intensity states as is observed in many HMXBs. \ros pointed 
observations covering a full binary cycle of $\sim$4~days depicted the source 
in a low state. Two short X-ray flares reaching peak luminosities close to 
$\rm 10^{38}\ erg\ s^{-1}$ and one long duration flare were recorded. The 
low-state spectrum has been found to be softer than the reported high-state 
spectrum. SMC~X-2 is a Be type transient discovered already 14~years before 
the \ros observations but has only been seen once in outburst. A second 
outburst has been detected by \ros during a very short pointing. The soft 
X-ray spectrum of the source has been measured for the first time. A large 
absorbing column has been found. RX~J0051.8-7231, a possible HMXB, has been 
discovered in \ein observations as a variable source and has been extensively 
studied with \ros . RX~J0101.0-7206 has been discovered during an X-ray 
outburst at the north-eastern boundary of the giant SMC HII region N66 with a 
luminosity of $\rm \sim10^{36}\ erg\ s^{-1}$. RX~J0049.1-7250 has been 
discovered with \ros . It is located north-east of the SMC supernova remnant 
N~19. It shows high absorption which indicates that the source is behind the 
SMC and has been detected with most probably a luminosity at the Eddington 
limit during the X-ray bright state. No optical counterpart has been found. 
The luminosity distribution deduced for the SMC X-ray binaries matches within 
the small number of the objects considered the distribution of galactic bulge 
sources. Scaling with the mass ratio between the SMC and our Galaxy of 
$\rm {1\over{60}}$ (McGee \& Hindman 1971) $\sim$400 hard X-ray binaries and 
$\sim$250 supersoft X-ray binaries would be expected to be seen in our Galaxy 
in case the star formation rate in the SMC and in our Galaxy is similar (cf. 
for a discussion Clark et al. 1978). As the metallicity of the SMC is strongly
reduced one can assume that there are not many hidden sources and this 
extrapolation is close to the total number of active hard and supersoft X-ray 
sources. The number of 250 supersoft sources would be reasonably close to the 
number of $\sim$1000 expected galactic supersoft sources (Rappaport et al. 
1994). van den Heuvel (1992) estimated the number of HMXBs in the Galaxy 
containing a NS to $\rm \sim10^4$. Obviously only a small fraction of them is 
active with luminosities above a few $\rm 10^{35}\ erg\ s^{-1}$.

\acknowledgements
P. Kahabka is a EC Human Capital and Mobility fellow under contract NR. 
The \ros project is supported by the Max-Planck-Gesellschaft and 
the Bundesministerium f\"ur Forschung und Technologie (BMFT). This research 
made use of the Simbad data base operated at CDS, Strasbourg, France. Part 
of the work has been performed during the stay of P.Kahabka at the 
Max-Planck-Institut f\"ur extraterrestrische Physik in Garching. We thank 
Ed van den Heuvel and Jan van Paradijs for reading the manuscript. We thank 
the referee Manfred Pakull for his detailed comments, which helped to improve 
the manuscript considerably. 


\end{document}